\documentclass[graybox]{svmult}

\usepackage{xcolor,soul} 
\usepackage{type1cm}        
\usepackage{makeidx}         
\usepackage{textgreek}
\usepackage{graphicx}        
\usepackage{multicol}        
\usepackage[bottom]{footmisc}
\usepackage{amsmath}
\usepackage[style=phys,biblabel=brackets, url=true]{biblatex}
\usepackage{newtxtext}       %
\usepackage[varvw]{newtxmath}       
\usepackage{caption,subcaption}
\usepackage{booktabs}
\usepackage[T1]{fontenc}

\makeatletter
\renewcommand{\p@subtable}{}%
\makeatother

\DeclareCaptionLabelFormat{customsub}{(#2)}
\makeindex             

\begin{document}

\title*{Can machines learn density functionals?  Past, present, and future of ML in DFT}
\author{Ryosuke Akashi, Mihira Sogal and Kieron Burke}

\institute{Ryosuke Akashi \at National Institutes for Quantum Science and Technology, 2-12-1 Ookayama, Meguro-ku, Tokyo, Japan, \email{akashi.ryosuke@qst.go.jp}
\and Mihira Sogal and Kieron Burke \at University of California, Irvine, CA 92697, USA
\\
\and March 3, 2025, submission to \textit{Machine Learning in Condensed Matter Physics - Significance, Challenges, and Future Directions} (Springer Series in Solid-State Sciences)}

%
%
\maketitle

\abstract{\\
Density functional theory has become the world's favorite electronic structure method, and is routinely applied to both materials and molecules.  Here, we review recent attempts to use modern machine-learning to improve density functional approximations. 
Many different researchers have tried many different approaches, but some common themes and lessons have emerged.  We discuss these trends and where they might bring us in the future.}

\tableofcontents
\clearpage
\section{Introduction}

\setstcolor{blue}
Machine learning (ML) allows us to develop a calculable and reusable surrogate of any task comprising of regression (including classification or pattern recognition). After conquering human pattern recognition~\cite{Krizhevsky_NIPS2012} and some famous games~\cite{Mnih_arxiv2013, Silver_Nature2016}, this technology is rapidly percolating into every branch of quantum chemistry and physics as a provider of versatile tools. In these fields a general aim is to solve the electronic many-body Schr\"odinger equation. The human process of solving the Schr\"odinger equation is as follows: first, relevant degrees of freedom must be defined for the system, e.g. spatial and spin coordinates of atoms and electrons; next, the Hamiltonian corresponding to the system can be written explicitly. Finally, the Hamiltonian is diagonalized to find the wavefunctions (eigenvectors) and corresponding energies (eigenvalues) of the system. The exact numerical solution of the Schr\"odinger equation is extremely costly, scaling exponentially in the number of electrons in the system. As a result, exact diagonalization of the Hamiltonian is seldom computationally feasible (for realistic Hamiltonians), and approximate methods must be used instead. 

The most widely used method, and the focus of this Chapter, is density functional theory (DFT), for which Walter Kohn shared the Nobel Prize in Chemistry in 1998 ~\cite{Hohenberg-Kohn1964,Kohn-Sham1965, 1998NobelChemistry}. DFT is an alternative formulation of the many-electron problem in which the charge density $n(\mathbf{r})$ is treated as the fundamental object of interest as opposed to the wavefunction $\psi$. While $\psi$ is a function of the spatial and spin coordinates of every electron in the system (for a total of $4 M$ variables where $M$ is the number of electrons), $n(\mathbf{r})$ depends only on the three general spatial coordinates $(x,y,z)$. Thus, the framework of DFT dramatically speeds up the process of solving the many-electron problem. While DFT provides a strong balance between accuracy and computational cost, the cost of executing a DFT calculation is still significant, of order $O(M^3)$. Furthermore, even within DFT, approximations are required so that calculations can be executed at reasonable cost. Machine learning is a promising strategy by which to improve both the speed and accuracy of DFT calculations. In fact, the capacity for ML to accelerate the approximate solution of the Schr\"odinger equation spans the entire process outlined above, and is not restricted to DFT alone ~\cite{Lilienfeld2023_review}. Ongoing research even includes the development of ML methods that can bypass DFT altogether, such as machine-learned potentials trained on DFT data (see other chapters). Nonetheless, the contributions of ML to the advancement of DFT methods is a conceptually rich and a practically significant topic, and the following sections are devoted to a review of the recent progress. 

In Section 2 we explain the basic foundations of DFT, and introduce the problem of approximating the exchange-correlation functional. In Section 3 we showcase some historical developments of approximate functionals found by humans in order to contextualize the more recent developments from ML. We summarize in Section 4 the technical and theoretical achievements enabled by ML. Section 5 is devoted to discussions on the generalization accuracy of the machines seen in the literature and potential directions for furthur research. Section 6 concludes the review with some perspectives on the future.

\section{Background: Almost a century of DFT}

Density functional theory begins with the foundational papers of Thomas and Fermi ~\cite{Thomas_1927} ~\cite{Fermi_1928}.  Here we summarize modern DFT.

\subsection{Foundations}
The electronic $N$-body dynamics is governed by the following Hamiltonian
\begin{eqnarray}
H=\sum_{i=1}^{N}\left[-\frac{\nabla^{2}_{i}}{2}+V_{\rm ion}({\bf r}_{i})\right]+\frac{1}{2}\sum_{i\neq j}^{N}\frac{1}{|{\bf r}_{i}-{\bf r}_{j}|},
\label{eq:Hamiltonian}
\end{eqnarray}
and the Schr\"odinger equation for the $N-$body wave function $\Psi$ is
\begin{eqnarray}
H\Psi({\bf r}_{1},{\bf r}_{2},\cdots, {\bf r}_{N})=E\Psi({\bf r}_{1},{\bf r}_{2},\cdots, {\bf r}_{N}). 
\label{eq:Schroedinger-eq}
\end{eqnarray}
where $\mathbf{r}_i$ are the spatial coordinates of the electrons, and $\Psi$ is antisymmetric with respect to particle exchange. \\
In this chapter, we restrict ourselves to the non-relativistic, non-magnetic case and adopt the atomic unit ($e=c=\hbar=m_{\rm e}=1$). We also suppress spin degrees of freedom in all equations for simplicity--although these are included in all modern DFT calculations.

This equation determines the series of many-body eigenstates once the one-body potential $V_{\rm ion}({\bf r})$ generated by the ionic configuration is given. In this sense, the eigenstates and their energy ordering are well-defined functionals of $V_{\rm ion}(\mathbf{r})$, a scalar value on real space.

Hohenberg and Kohn formulated two important theorems~\cite{Hohenberg-Kohn1964}. First, the charge density distribution of the ground state $n({\bf r})$ determines the corresponding $V_{\rm ion}({\bf r})$ unambiguously (if it exists) that has such an $n({\bf r})$ in the ground state (in the non-degenerate case). Second, there exists a variational equation with $n({\bf r})$ as the variable that yields the ground state energy: Formally, there is a functional of the whole distribution of $n(\mathbf{r})$ bounded by the ground state energy for given $V_{\rm ion}({\bf r})$.
\begin{eqnarray}
    E[n,V_{\rm ion}]
    \equiv
    F[n] + \int d{\bf r}n({\bf r})V_{\rm ion}({\bf r})\geq E_{\rm GS}[V_{\rm ion}]
    ,
    \label{eq:var-energy}
\end{eqnarray}
for which the equality holds only when $n({\bf r})$ is equal to that at the true ground state for $V_{\rm ion}(\bf r)$. The functional $F[n]$, as expressed later by Levy~\cite{Levy1979} and Lieb ~\cite{Lieb-1983}, is
\begin{eqnarray}
    F[n]\equiv \min_{\Psi\rightarrow n}\langle \Psi|\left(\hat{T}_{\rm e}+\hat{U}_{\rm ee}\right)|\Psi\rangle,
    \\
    \hat{T}_{\rm e}=\sum_{i=1}^{N}\left(-\frac{\nabla^2_i}{2}\right), \ \ \hat{U}_{\rm ee}=\frac{1}{2}\sum_{i\neq j}\frac{1}{|{\bf r}_{i}-{\bf r}_{j}|},
\end{eqnarray}
where the minimization is taken with constraint that the many-body state $\Psi$ yields the given charge density distribution through
\begin{eqnarray}
    n({\bf r})=\int d{\bf r}_{2}d{\bf r}_{3}\cdots d{\bf r}_{N}|\Psi({\bf r},{\bf r}_{2},\cdots,{\bf r}_{N})|^2
    .
\end{eqnarray}
We express the dependence as functional by squared parentheses ``$[]$". The functional $F[n]$ is universal, which simply means it is a density functional with no dependence on $V_{\rm ion}(\mathbf{r})$.

The Euler equation for the ground state charge density is:
\begin{eqnarray}
    \frac{\delta F[n]}{\delta n({\bf r})}+V_{\rm ion}({\bf r})= 0
    ,
    \label{eq:variational}
\end{eqnarray}
if we fix the particle number. This equation clarifies the input-output relation of the ground-state quantum theory: Scalar function of space $(V_{\rm ion}({\bf r}))$ in, scalar function of space $(n({\bf r}))$ out. Of course this equation does not simplify the problem since, to obtain the exact formula of $F[n]$, we are required to solve the Schr\"{o}dinger equation. Thus, accurate and calculable models for $F[n]$ are essential for efficient calculations.

People have sought two paths for solving the DFT variational equation~(\ref{eq:variational}). The first is the Kohn-Sham (KS) framework, and the second is the older orbital-free (OF) DFT. For both, the derivation of useful approximations to $F[n]$ has been the challenge. The latter in principle yields cheaper computational workflows but simultaneously imposes the additional subordinate (and difficult) challenge of how to calculate the kinetic energy of without the concept of orbitals.

\subsection{The Kohn-Sham equations}
Kohn and Sham~\cite{Kohn-Sham1965} created a reference system that is formally solved more easily than the many-body Schr\"odinger equation, but yields the same ground-state charge density, via a transformation of the variational equation (\ref{eq:variational}), as
\begin{eqnarray}
&& \left[
-\frac{\nabla^2}{2}
+
V_{\rm eff}[n]({\bf r})
\right]
\varphi_{a}({\bf r})
=\varepsilon_{a}\varphi_{a}({\bf r}),\ \\ 
&&
n({\bf r})
=
\sum_{a}\theta(\mu-\varepsilon_{a})|\varphi_{a}({\bf r})|^2
\equiv
\sum_{a:{\rm occ.}}|\varphi_{a}({\bf r})|^2
.
\label{eq:KS-eq}
\end{eqnarray}
These are the Kohn-Sham (KS) equations. The effective potential $V_{\rm eff}[n]({\bf r})$ is related to the universal functional $F[n]$ through
\begin{eqnarray}
    V_{\rm eff}[n]({\bf r})
    =
    V_{\rm ion}({\bf r})
    +
    V_{\rm H}[n]({\bf r})
    +
    V_{\rm xc}[n]({\bf r}),
\end{eqnarray}
where
\begin{eqnarray}
   V_{\rm H}[n]({\bf r})=\int d{\bf r}' \frac{n({\bf r}')}{|{\bf r}-{\bf r}'|}, \ \
   V_{\rm xc}[n]({\bf r}) = \frac{\delta E_{\rm xc}}{\delta n({\bf r})},
\end{eqnarray} and the exchange-correlation (XC) energy functional is defined as:
\begin{eqnarray}
    E_{\rm xc}[n]
    =
    F[n]-T_{\rm s}[n]-E_{\rm H}[n],
\end{eqnarray}
where
\begin{eqnarray}
    T_{\rm s}[n] = \min_{\Phi\rightarrow n} \langle \Phi |\hat{T}_{\rm e}|\Phi\rangle, \ \ 
    E_{\rm H}[n] = \frac{1}{2}\int d{\bf r}d{\bf r}' \frac{n({\bf r})n({\bf r}')}{|{\bf r}-{\bf r}'|}.
\end{eqnarray}
The kinetic energy functional $T_{\rm s}[n]$ is defined by the constrained minimization of Eq. (4). The majority of kinetic energy is thus treated by an auxiliary set of single-particle orbitals $\{\varphi_{a}\}$ ($a$: spin-orbital) that form $\Phi$, while the remaining quantum effects are condensed into the exchange-correlation energy $E_{\rm xc}[n]$. The ground state total energy $E_{\rm GS}[V_{\rm ion}]$ is derived from the stationary solution of the KS equations by
\begin{eqnarray}
    E_{\rm GS}[V_{\rm ion}]
    =
    \sum_{a:{\rm occ.}}\varepsilon_{a}
    -
    E_{\rm H}[n]+E_{\rm xc}[n]-\int d{\bf r}n({\bf r})V_{\rm xc}({\bf r}).
\end{eqnarray}

This framework opens a path to practical first-principles calculations for the total energy of electronic systems such as molecules and solids, if any useful approximations to $E_{\rm xc}$ are available. It should be considered as a formalization of Slater's $X_\alpha$ method ~\cite{SlaterXAlpha}. The development of useful models for $E_{\rm xc}[n]$ has thus become a grand challenge in theoretical chemistry, condensed matter physics, and material science.

The reference one-particle system can be taken somewhat arbitrarily, as we can formulate an equation that is different from the original KS but in principle gives an identical ground state charge density~\cite{Seidl-GKS1996}. For some arbitrary operator $\hat{O}$ on the KS orbitals, we can construct the following system:
\begin{eqnarray}
    \left[
    \hat{O}[\{\varphi\}]
+V^{\rm gKS}_{\rm eff}[n]({\bf r})
\right]
\varphi_{a}({\bf r})
=\varepsilon_{a}\varphi_{a}({\bf r})
,
\end{eqnarray}
\begin{eqnarray}
&&
\frac{\delta S[\Phi]}{\delta \varphi_{a}({\bf r})}
\equiv
\hat{O}[\{\varphi\}]\varphi_{a}({\bf r}) 
\label{eq:S-diff}
\\
&&
V^{\rm gKS}_{\rm eff}[n]({\bf r})
\equiv V_{\rm ion}({\bf r})+\frac{\delta \left(F[n]-F^{\rm s}[n]\right)}{\delta n({\bf r})}, \ \ 
F^{\rm s}[n]
\equiv \min_{\Phi \rightarrow n}S[\Phi]
.
\end{eqnarray}
Here, $S[\Phi]$ is an arbitrary functional of the Slater state $\Phi$ with subtle mathematical requirements to $F^{\rm s}[n]$. This \textit{generalized} KS framework justifies including supplemental one-body operators (such as the Fock exchange) in the KS equations . This generalization can make ground-state calculations much more convenient while yielding practically identical ground-state energies. However, the spectrum of eigenvalues can be much improved in a Generalized KS calculation ~\cite{KronikGKS}.

\subsection{Orbital-free DFT}
 Orbital-free DFT (OF-DFT) aims to solve the variational equation~(\ref{eq:variational}) directly in terms of the density $n({\bf r})$. As in KS-DFT, a separation of the universal functional $F[n]$ is used; $F[n]=T_{\rm s}[n]+E_{\rm H}[n]+E_{\rm xc}[n]$. The OF-DFT approach poses additional challenges: (i) First, $T_{\rm s}[n]$ has to be modeled without solving for single-particle orbitals. (ii) Second, treatment of $V_{\rm ion}(\bf r)$ also raises an issue. In the KS approach one can omit the core electrons from the equations by using a pseudopotential to describe the effects of the core states on the valence states by the valence-orbital dependent projector (this becomes problematic in OF-DFT). We do not discuss pseudopotentials in depth here, but refer the reader to Ref.~\cite{Mi_ChemRev2023} for a thorough review.

\section{Human development of the functional}
Various conditions on the exact functional are well known~\cite{Sun_PRL2015}. The early development of approximate functionals was often grounded in interpolations between these well-studied exact conditions. Approximations constructed this way automatically satisfy exact conditions--but are also calculable in the intermediate regimes.  Machine learning brings  a sophisticated tool to the interpolation step. We review a few approximate functionals so that the historical context of the more recent ML development may be clarified.

Conventionally the XC energy models have been developed based on the separation into the exchange and correlation parts $E_{\rm xc}[n]=E_{\rm x}[n]+E_{\rm c}[n]$. The exchange part is defined by the Coulomb exchange integral with an appropriate single-particle basis. Defined with the KS orbitals, it is functional of $n(\mathbf{r})$, as each orbital is an implicit density functional (here spin is required, as there is no exchange contribution from opposite spins):
\begin{eqnarray}
E_{\rm x}[n]
=
-\frac{1}{2}\sum_{a,b:{\rm occ.}}\int d{\bf r}d{\bf r}'\frac{\varphi^{\ast}_{a}({\bf r})\varphi^{\ast}_{b}({\bf r}')\varphi_{b}({\bf r})\varphi_{a}({\bf r}')}{|{\bf r}-{\bf r}'|}
.
\end{eqnarray}
The correlation part $E_{\rm c}[n]$ contains all  other quantum effects. $E_{\rm x}[n]$ and $E_{\rm c}[n]$ can each be written as an integral over an XC energy per electron, e.g:
\begin{eqnarray} 
    E_{\rm c}[n]
    =
    \int d{\bf r} n({\bf r})\varepsilon_{\rm c}[n]({\bf r}).
\end{eqnarray}
Note that choices of energy per electron are ambiguous since they can have arbitrary components that integrate to zero in total ~\cite{Burke_JCP1998}. The definitions above can also be reasonably extended for the two-component spin DFT, where the functionals depend on $n_{\uparrow}(\mathbf{r})$ and $n_{\downarrow}(\mathbf{r})$ representing the spin-up and down components of the electron charge density. The relative spin polarization introduced $\zeta({\bf r})=(n_{\uparrow}({\bf r})-n_{\downarrow}({\bf r}))/(n_{\uparrow}({\bf r})+n_{\downarrow}({\bf r}))$ is often used to parameterize spin dependence.

\subsection{Jacob's Ladder of Density Functional Approximations}
Approximations to the XC functional can be organized into a hierarchy which Schmidt and Perdew ~\cite{Jacobs_ladder} coined as Jacob's ladder. The ladder spans the distance between the (crude) Hartree approximation and the sought-after Heaven of Chemical Accuracy (the $\sim 1$ kcal/mol accuracy required to predict the energy of a chemical reaction). The lowest ``rung" of the ladder is the simplest approximation to the XC functional: the local density approximation (LDA). In the LDA, the XC functional is constructed as a \textit{local} functional of the density--that is, it depends on the density at that point alone. Improvements on the LDA (the higher ``rungs") are guided by the concept of a gradient expansion ~\cite{Hohenberg-Kohn1964,Kohn-Sham1965}. Going up the rungs, the approximate models of the exchange-correlation energy densities incorporate dependence on the charge density at the site $n({\bf r})$ [local (spin) density approximation, L(S)DA], its gradient $\nabla n({\bf r})$ (generalized gradient approximation, GGA), Laplacian and/or kinetic energy density (meta GGA), occupied KS orbitals, (hybrid), and unoccupied ones (double hybrid, RPA). Below we review some example functionals from the various rungs. An excellent summary of approximate XC functionals can be found in ~\cite{ToulouseBook}.
 
\subsection{Local density approximation}
In the LDA, the energies per electron are given by
\begin{eqnarray}
    \varepsilon^{\rm LDA}_{\rm x, c}[n]({\bf r})
    =
    \varepsilon^{\rm unif}_{\rm x, c}(n({\bf r}))
    .
    \label{eq:LDA}
\end{eqnarray}
where $\varepsilon^{\rm unif}_{\rm x, c}(n({\bf r}))$ is the energy per electron of the uniform gas with density $n$.
To develop the model for $\varepsilon_{\rm x, c}(n)$, people refer to the system’s behavior in the uniform electron gas limit $V_{\rm ion}=0$. Often the charge density is characterized by the Wigner-Seitz radius $r_{\rm s}= \sqrt[3]{3/(4\pi n)} \equiv r_{\rm s}(n)$. The dependence on the spin density is also included, called the local spin density approximation (LSDA):
\begin{eqnarray}
    \varepsilon^{\rm LSDA}_{\rm x, c}[n_{\uparrow},n_{\downarrow}]({\bf r})
    =
    \varepsilon^{\rm unif}_{\rm x, c}(n({\bf r}), \zeta({\bf r})).
\end{eqnarray}
The exchange energy per electron in the uniform electron gas is analytically calculated to be
\begin{eqnarray}
    \varepsilon^{\rm unif}_{\rm x}(n)
    =-\frac{3}{4}\left(\frac{3}{\pi}\right)^{1/3}n^{1/3}
    .
    \label{eq:exchange-UEG}
\end{eqnarray}
First Bloch ~\cite{Bloch1929}, then Dirac~\cite{Dirac_1930} derived this  form for Eq.~(\ref{eq:LDA}).

\subsubsection{Correlation of the Uniform Gas}
Several parameterizations of $\varepsilon^{\rm unif}_{\rm c}(n)$ exist, the most commonly used being VWN ~\cite{VWN1980}, Perdew-Zunger ~\cite{PZ81}, and PW92 ~\cite{PW92}. All three were constructed with the same basic strategy. First, an analytical form is chosen such that it recovers the correct asymptotic behavior in either the high-density ($r_s \to 0$) or low-density ($r_s \to \infty$) limit, or both. This is followed by numerically fitting a small number of free parameters to the Monte Carlo results of Ceperley and Alder\cite{Ceperley1978,Ceperley-Alder1980}; the interpolation serves to extend the range of $r_s$ for which the model is accurate. All three of these parameterizations agree with one another to within a roughly 2-percent error (which is substantially smaller than the error of LDA for inhomogenous systems). The parameter determination for the spin-polarized case used Misawa's spin scaling for the correlation energy ~\cite{Misawa1965}.

\subsection{Generalized gradient approximation (GGA)}
The next rung of Jacob's ladder corresponds to inclusion of the density gradient
\begin{eqnarray}
    \varepsilon^{\rm GGA}_{\rm x c}[n]({\bf r})
    =
    \varepsilon^{\rm GGA}_{\rm x c}(n({\bf r}), |\nabla n({\bf r})|),
\end{eqnarray}
called generalized gradient approximation (GGA). The development of useful GGA model functionals has been much advanced through '80s and '90s. Although there have been many proposed models, we here explain the Perdew-Burke-Ernzerhof (PBE) model~\cite{GGAPBE1996}, that well represents how the analytical human functional development proceeds.

The PBE correlation energy per electron is
\begin{eqnarray}
    \varepsilon^{\rm PBE}_{\rm c}[n_{\uparrow},n_{\downarrow}]({\bf r})
    = \varepsilon^{\rm unif}_{\rm c}(r_{\rm s}({\bf r}),\zeta({\bf r})) + H(r_{\rm s}({\bf r}),\zeta({\bf r}),t({\bf r})),
    \label{eq:GGA-correlation}
\end{eqnarray}
where $t$ is a dimensionless measure of the gradient. The form of $H$ is chosen to satisfy three conditions: (a) In the small-$t$ limit it reduces to the second order gradient expansion~\cite{Wang-Perdew1991} $H \sim \beta t^2$, from which $\beta$ is known to be $\simeq 0.066725$ . (b) In the rapidly varying limit $t\rightarrow \infty$, the correlation energy should vanish: $H \rightarrow -\varepsilon_{\rm c}^{\rm UEG}$. (c) The total correlation energy must scale to a constant under uniform scaling to the high-density limit (Ref.\cite{Levy-scaling1989}).

The exchange energy per electron in the unpolarized case $\zeta=0$ is formulated as
\begin{eqnarray}
    \varepsilon^{\rm PBE}_{\rm x}[n]({\bf r})
    \equiv \varepsilon_{\rm x}^{\rm unif}(n({\bf r}))F_{\rm x}(s({\bf r})),
    \label{eq:GGA-exchange}
\end{eqnarray}
with $s=|\nabla n|/(2k_{\rm F}n)$ and $F_{\rm x}$ is called the \textit{enhancement factor}.
The spin dependent formula is derived by the exact spin-scaling relationship for exchange ~\cite{Oliver-Perdew-spin-scaling}. The PBE form satisfies the following three conditions: (d) Under the uniform density scaling described above, $E_{\rm x}$ must scale linearly (Ref.~\cite{Levy-Perdew-exchange-scaling1985}). (e) It cancels the gradient correction to the correlation energy as $s\rightarrow 0$ so that the model recovers the accurate linear response of LDA. (f) It satisfies the Lieb-Oxford inequality~\cite{Lieb-Oxford-bound} for all $n$ and $s$, thus ensuring satisfaction for all possible densities.

In the above construction, the formulas have four parameters determined from the conditions. The artificial aspect is in selecting the forms of $H$ [Eq.~(\ref{eq:GGA-correlation})] and $F$ [Eq.~(\ref{eq:GGA-exchange})], which are mathematically simple for humans and minimally parametrized. Many variations on the PBE approximation have since been suggested, including RPBE ~\cite{RPBE1999}, revPBE~\cite{revPBE1998}, and PBEsol~\cite{PBEsol2007}.

\subsection{Physically fitted functionals}
\label{sec:semiempirical}
Becke initiated a complementary approach for functional development. He introduced a model exchange energy within the GGA with a single parameter and determined it so that it optimally reproduces the Hartree-Fock energies in 6 rare gas atoms up to Rn~\cite{Becke_PRA1988}, called B88. This parametrization satisfied exact conditions on spin-scaling and recovery of the uniform limit. Moreover, the form of the exchange energy per electron yields the correct asymptotic behavior in the evanescent region of an atom or molecule--a crucial property of chemical bonding ~\cite{ElliorBurkeB88},~\cite{CancioSquarePeg}.

\subsubsection{Hybrid Functionals}
Based on intuitive ideas of the adiabatic connection curves of approximate functionals (see also Ref.\cite{Teale_JCTC2025} for recent advance), Becke further proposed~\cite{Becke_JCP1993_3} to modify his B88 model by replacing a fraction of LSDA exchange with exact exchange, and adding correlation at the GGA level:
\begin{eqnarray}
    E_{\rm xc}^{\rm B3PW91}=E_{\rm xc}^{\rm LSDA} + a_{0}(E_{\rm x}-E_{\rm x}^{\rm LSDA})+a_{\rm x}(E_{\rm x}^{\rm B88}-E_{\rm x}^{\rm LSDA})+a_{\rm c}(E_{\rm c}^{\rm PW91}-E_{\rm c}^{\rm LSDA}),\nonumber \\
    \label{eq:Becke_formula}
\end{eqnarray}
where $E_{\rm c}^{\rm PW91}$ is from Ref.~\cite{Perdew_book1991}. He chose the three parameters $(a_{0}, a_{x}, a_{c})$ so that experimentally observed reaction energies for approximately 100 reactions are well reproduced. The variant of this model with $E_{\rm c}^{\rm PW91}$ replaced by the Lee-Yang-Parr form~\cite{Lee_PRB1988} is the B3LYP functional~\cite{Kim_JPC1994,Stephens_JPC1994}, the most used hybrid functional (4th rung of the ladder) in quantum chemistry. Although Becke himself emphasized the physics~\cite{Becke_JCP1993_3}, the scheme stimulated researchers, yielding various data-tuned functionals in the next decades (see Ref.~\cite{Mardirossian_MolPhys2017} for a review). Many of these developments, not originally declaring themselves as ML, used the same fundamental strategies as modern ML-DFT; but often were contrary in spirit to Becke's philosophy.

\section{Machine-learning}
Machine learning technology has introduced a paradigm that is in some ways similar to the variational theory for quantum many-body Hamiltonian. When we calculate the many-body ground-state wavefunction of an interacting quantum system by the variational method, we prepare a model wavefunction with numerous parameters and optimize those so that the energy expectation value is minimized. The derived wavefunction is used for calculating the observable properties such as correlation functions, but of course this specific wavefunction is only used for this one system.

This scheme is not regarded as useful for finding approximate functionals in DFT since the latter aims to utilize the derived functional for systems not referenced in the optimization step. Even if we could achieve an extreme accuracy for a specific system, such an optimized model would never be applicable to others. To this field, ML has provided a framework to optimize a model function for {\it a group} of systems, keeping its accuracy even when applied out of the group (via some form of regularization). The surge of ML developments in DFT has demonstrated that highly parameterized functionals may have general accuracy, at least for an appropriately defined range of applications.

In this section, we attempt to summarize numerous recent results applying  ML to design density functionals, and use these to draw general insights into the possibilities that ML has opened for improved DFT calculations. 

\subsection{General discussion}
ML in DFT generally concerns modeling of the functionals appearing in the previous sections, such as exchange-correlation energy $E_{\rm xc}[n]$. The tricky thing with DFT is that the density functionals and $n(\mathbf{r})$ are, in practice, both outputs of the Hamiltonian that are simultaneously determined after solving the KS equations. Because of this there has been no convenient analytic theory that relates $n(\mathbf{r})$ and functionals as input and output, unlike the wave-function based perturbation theory that relates the single-particle basis to physical quantities. Machine learning provides versatile tools to optimize models relating those using a large number of pairs $(n(\mathbf{r}), {\rm functional})$ produced by accurate calculations in specific cases.

The typical procedure starts from modeling of a target functional in DFT, dubbed as $f[n]$, by any regressor that can accept any value of $n(\mathbf{r})$. Practically this is implemented as a vector-to-scalar\footnote{Hereafter we assume $f$ to be a scalar for simplicity but we can also take $f$ as vector or tensor.} mapping $n (\mathbf{r})\rightarrow f[n]$, where $n(\mathbf{r})$, a non-negative distribution in space, is approximately represented as a vector; $n(\mathbf{r})\approx{\bf n}$. How to efficiently represent ${\bf n}$ is also of a matter of ongoing development. The regressor model chosen has many parameters. We then collect specific examples of values $(n^{(i)}, f^{(i)}[n])$ for various different potentials to serve as training data. These reference data must be accurate enough for the intended applications of the ML model. The optimization of the model parameters is executed so that a loss function
\begin{eqnarray}
    L=\frac{1}{N_{\rm data}}\sum_{i=1}^{N_{\rm data}}D(f^{(i)},f_{\rm model}[n^{(i)}])+({\rm reg.}).
\end{eqnarray}
is minimized. Here $D(f,f')$ represents any non-negative function that measures distance between inputs $f$ and $f'$; e.g., $D(f,f')=(f-f')^2$ and ``reg." denotes the regularization term that prevents overfitting to the training data. While the building blocks for this procedure predate ML,  ML technology has provided both powerful model functions and efficient methods for  optimization and regularization. 

\subsubsection{Universal regressors}
In the DFT context, two approaches are often used for the above $f_{\rm model}$: Kernel ridge regression (KRR) and neural network (NN). We explain the features of those models very briefly. 

Kernel ridge regression (KRR)~\cite{Bishop_PRML} has a form
\begin{eqnarray}
f({\bf v}_{\ast})=\sum_{j=1}^{N_{\rm data}}\alpha_{j}k({\bf v}_{j},{\bf v})
,
\end{eqnarray}
where ${\bf v}$ denotes any descriptor vector for $f$. The kernel function $k({\bf x},{\bf x}')$ can have any form as long as the $N_{\rm data}\times N_{\rm data}$ matrix formed by any set of data $\{{\bf v}_{1},{\bf v}_{2},\cdots {\bf v}_{N_{\rm data}}\}$, $\mathcal{K}_{ij}=k({\bf v}_{i},{\bf v}_{j})$, is positive definite. Assuming ridge regularization, the weight $\alpha_{i}$ is exactly determined by a matrix inversion of size $N_{\rm data}\times N_{\rm data}$.

Neural networks (NN) come in many varieties. The typical fully connected NN is formed by repetition of non-linear vector-to-vector transformation ${\bf v}\rightarrow{\bf v}'$
\begin{eqnarray}
{\bf v}' = \phi(W{\bf v}+{\bf b})
,
\end{eqnarray}
which is often represented by the celebrated graph (probably seen in other Chapters) composed of layers of nodes and edges connecting those. Here, $W$ and ${\bf b}$ are the parameter matrix and vector to be optimized and $\phi$ denotes some non-linear transformation operated on the respective vector components. The intermediate vector dimensions are arbitrary.

Both KRR and NN are both assured to be universal approximators~\cite{Micchelli_JMLR2006,Hornik_NN1991,Cybenko_MathCSS1989, Kolmogorov1956}, in the sense that the model can numerically reproduce \textit{any} reference function--provided that the model has a sufficient number of parameters. This is achieved by increasing the reference data and intermediate dimensions for the KRR and NN, respectively. 

The universal character of these models is quite beneficial for developing approximate density functionals. In principle, the functionals $f[n]$ have rigorous (but not directly calculable) forms that are only inferred from specific cases. NN and KRR-based models are capable of representing any dependences between $n(\mathbf{r})$ and $f$, by fitting the specific data. They can also attain some robustness against overfitting when the optimized models are applied to unseen cases.

Although those two approaches are mathematically equivalent in some limits~\cite{Lee2018_arxiv}, a crucial difference between the KRR and NN exists in practice. KRR needs to store the training data for later calculations, but once the definition of the loss and training data are specified, the weights $\{\alpha_{i}\}$ can be rigorously determined. The NN, on the other hand, treats the training data explicitly only at the model optimization. The information of the training systems are then encoded implicitly in the NN model parameters. However, the optimum parameters depend on the optimization methods and, if one adopts stochastic methods such as dropout and batch optimization, their specific values are not reproducible even with fixed NN architecture and training systems. This is quite disturbing from a traditional quantum-chemical viewpoint.

\begin{figure}[t]
\sidecaption
\includegraphics[scale=.25]{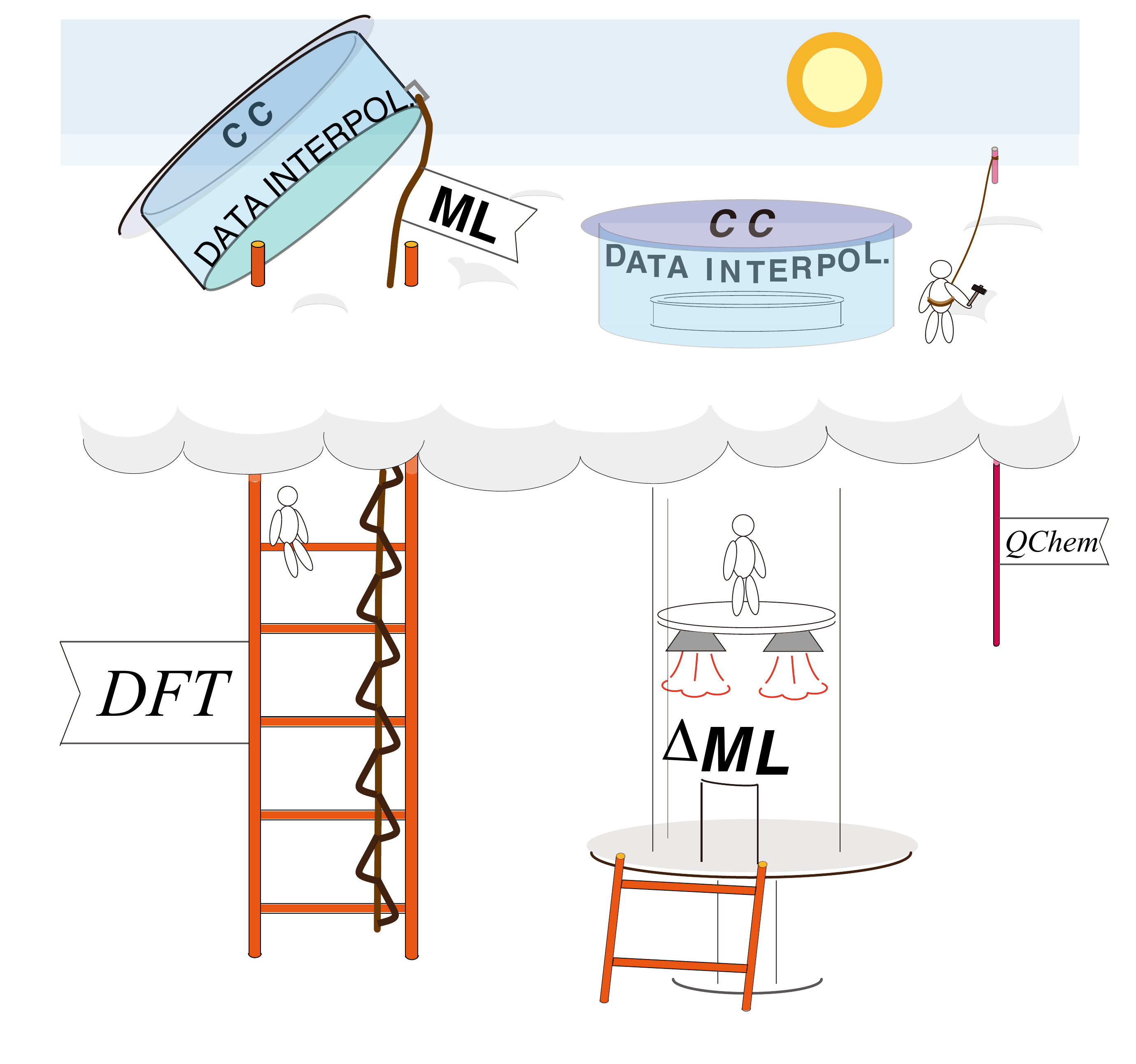}
\caption{Illustration of the relation between traditional DF and quantum chemical theories and ML-based DFT developments. Machine learning can help people climb (with less computational cost) toward the chemical accuracy heaven by supplementing supporting rungs (=variables) along with DFT. A more convenient path by ML is available that boosts people from some low rung of Jacob's ladder to the heaven. Either way people can sightsee the heaven, as far as the reference data and quantum chemical theory allow. Researchers yet need to climb the traditional theoretical paths of quantum chemistry and DFT without ML for working on the ML limitation.}
\label{fig:Ladder_boost}       
\end{figure}

\subsubsection{Limitation by data}
The accuracy of ML-based functionals is limited by the training data, but accurate data available now for training is still little in amount. The most accurate data are those from experimental observations, which are limited in number of the systems as well as in the variety of quantities. Few quantities are measurable with enough accuracy, like the energy-related quantities and structural properties such as bond lengths and angles and lattice parameters. Hence, in most of the ML-based DFT developments, researchers refer to databases augmented by theoretical calculations. In datasets based on existing theoretical methods, the procedure to generate the training data pair $(n(\mathbf{r}), f)$ is free from any uncertainty like noise in experimental data--allowing for more consistent benchmarking and comparison across ML models.  Currently, the most accurate and executable solver for molecules is the coupled cluster (CC) method~\cite{Bartlett_AnnRevPhysChem1981, Purvis_JCP1982}. Fortunately, in many cases coupled-cluster singles and doubles and perturbative triples (CCSD(T)~\cite{Raghavachari_CPL1989}) achieves chemical accuracy (errors less than 1 kcal/mol or about 30 meV for covalent bonds in small molecules) and therefore versatile ML-DFT functionals with CC accuracy are worth aiming for. 

\subsubsection{ML Usage with DFT}
People have explored various fitting tasks concerning the density functionals using ML. Theoretically there is a nuanced difference among the papers in the level of the ML usage: They either (i) preserve the rigorous framework of DFT, (ii) supplement or completely replace the variable $n({\bf r})$ with atomic potentials $V_{\rm ion}({\bf r})$, or (iii) with element labels. Level (ii) is distinct from (iii) as it is based on the potential functional theory, which takes $V_{\rm ion}({\bf r})$ as the variable for the corresponding ground state. This is a formal dual of DFT~\cite{Lieb-1983,Yang_PRL2004} and therefore is capable of carrying the same information. Using the element descriptors in addition to $n({\bf r})$ seems useful for efficient fitting in some cases, but at the price of departure from the density-potential functional framework. 

Across the studies we address in this review, ML methodologies fit the Jacob's Ladder paradigm to varying degrees. Some try better parametrizations of the functionals used in the KS or OF framework, adapting the definitions of variables in the ladder rungs. Others seek modified implementations of the ladder, by laying ``rungs" of different variables other than original, like the non-local weighted density~\cite{Gunnarsson_PRB1976}. Still others fit the difference of the functionals calculated with the high-level DF or quantum chemical approximations and low-level DFT by extra variables that may be of level (ii) or (iii). Such studies would be understood as building an ML path to Heaven on the basis of (low-level) DFT, among which we find the current most successful achievements in the cost-accuracy balance as we will highlight below. For all those, we again remind that the readers should be aware of the limitation by data. We sketch the situation by Fig.~\ref{fig:Ladder_boost}.

\subsubsection{Interpolation, extrapolation, and generalization}
In principle, a trained model should not be applied to systems that are not represented in the training data. For a trained machine the target systems are classified into two groups. Systems for which the model is expected to provide accurate results are called ``in-domain" or ``in-sample". Systems that are poorly represented by the training data, and therefore are not expected to be accurately captured by the model, are termed ``out-of-domain" or ``out-of-sample". In this review, we use ``interpolation" to refer to the application of a model to an in-domain system, and ``extrapolation" to refer to the application of a model to an out-of-domain system.

Note that whether a task constitutes interpolation or extrapolation  is judged only after application; the system for which the machine yields accurate predictions is {\it found to be} in the ``domain of interpolation". The interesting issue for ML applications to DFT would be the possibility that any training scheme can result in an interpolation domain which is broader than one would expect from the training data. Because within the DFT framework the variable is $n(\mathbf{r})$, the original atoms are anonymized, and therefore the interpolation domain can become broader than expected from the literal atomistic composition of the systems. We call the success of the models to unseen systems ``generalization", rather than extrapolation.

\subsection{Fitting of functionals}
Here we show some of the successful cases for making well-behaved density functionals $f[n]\approx f({\bf n})$ with machine learning. The functional $f[n]$ could be the total energy, the exchange-correlation energy, the exchange-correlation potential, forces, or any other quantities related to the ground state determined by the specified $n(\bf r)$.

\subsubsection{Kinetic energy}
In 2012, Snyder et al.~\cite{Snyder_PRL2012, Snyder_JCP2013} calculated the exact ground-state charge density and non-interacting kinetic energy in a one-dimensional model, where the model ionic potential was replaced by a sum of Gaussians controlled by a few parameters. They demonstrated that the kinetic energy $T_{\rm s}[n]$ is well fitted by the KRR model as a functional of $n(\mathbf{r})$ being a vector $\{n(x_{1}),n(x_{2}), \cdots\}$ on a discretized spatial grid. This work is the first demonstration that a fully nonlocal functional can be implemented for accurate reuse, as long as the range of application is well controlled: In this study the training and test potentials were both within the specific range of the Gaussian model parameter space. This has been also shown to hold for the exchange-correlation energy in one dimension~\cite{Li_Wagner_PRB2016}. 

The faithful fitting of fully nonlocal functionals in three dimensions is far more difficult since the representation of $[n]$ requires a large amount of data. Yao and Parkhill~\cite{Yao_Parkhill2016} pursued $T_{\rm s}[n]$ by a convolutional NN form, whereas Brockherde et al.~\cite{Brockherde_NatComm2017} demonstrated a bypass of the functional relation $V_{\rm ion}\rightarrow n$ in the KS framework by KRR for molecules in 3D, avoiding even the explicit construction of $T_{\rm s}[n]$.

Semilocal approaches to $T_{\rm s}[n]$ with ML have been widely explored. Seino et al.~\cite{Seino_JCP2018} targeted the kinetic energy density $\tau({\bf r})$ defined with a local decomposition $T_{\rm s}[n]=\int d{\bf r}\tau([n];{\bf r})$. They collected the grid data for the converged $n({\bf r})$ and $\tau({\bf r})$ calculated from KS-DFT for (i) atoms from H to Ne and (ii) 19 small molecules composed of C, H, O, and N, and separated those into the training and test grid points in dataset (i) and (ii), respectively. They implemented the semilocal form $\tau[n](\bf r)=\tau(n({\bf r}),|\nabla n({\bf r})|, \nabla^2 n({\bf r}),|\nabla \nabla^2 n({\bf r})|)$ using the NN and examined the fitting accuracy. Golub and Manzhos dug further into the fitting problem with focus on the instability near the atom~\cite{Golub_PCCP2019} and ML model dependency~\cite{Golub_CPC2020}. Semilocal modeling tends to suffer from error in molecular reaction energies of order 0.1 Hartree compared with the KS-DFT, which still remains a challenge even with ML.  Still, it may be useful for estimating equilibrium structural properties as shown by Imoto et al.~\cite{Imoto_PRR2021}. Inclusion of the atomic kind and positions as descriptors can be useful for improvement~\cite{Seino_CPL2019,Manzhos_JCP2020}, though is not a conclusive solution. All such attempts must account for the ambiguity in the energy density discussed earlier.

Nonlocal density descriptors are generally expressed in the following convolution
\begin{eqnarray}
    \varrho({\bf r})=\int d{\bf r}' w({\bf r}-{\bf r}')n({\bf r}')
    ,
    \label{eq:Nonlocal-n}
\end{eqnarray}
and seem useful for capturing the nonlocal nature of $T_{\rm s}[n]$. Note that the convolution weight $w$ can also be a target of ML modeling. Ryczko et al.~\cite{Ryczko_JCTC2022} demonstrated a fitting of a voxel-to-voxel NN model for $n({\bf r})$ to $\delta T_{\rm s}/\delta n({\bf r})$ for graphene sheet. The decomposition of $n({\bf r})$ by atom-centered basis functions has also been used as an efficient method for incorporating the nonlocality~\cite{Remme_JCP2023,Zhang_NatComputSci2024}. The nonlocal descriptors~\cite{Sun_Chen_PRB2024} are generally found to reduce the fitting error by an order of magnitude compared to semilocal approximations.


Using machine-learned energies for the variational solution Eq.(\ref{eq:variational}) imposes another challenge since an ML model fitted to  energies suffers from numerical noise when we take derivatives~\cite{Snyder_PRL2012}. Training sets can never be dense enough to infer the derivatives in all directions in the high dimensional descriptor space. The numerical instability induced by this may be mitigated by stopping the derivatives going out of the sampling space~\cite{Snyder_PRL2012,Snyder_IJQC2015,Li_JQC2016}. The finite difference method is also a stable choice~\cite{Golub_CPC2020}, but with increased computational cost relative to the direct derivatives. Defining the loss function to include the functional derivatives was also shown to be effective~\cite{Meyer_JCTC2020,Imoto_PRR2021}. For NN models, the noisy behavior may be mitigated by carefully designing the NN architecture~\cite{Costa_PRE2022}. 

The kinetic ``potential" used for the variational approach, $\delta T_{\rm s}[n]/\delta n({\bf r})$, has significant nonlocality. To fully take advantage of the OF framework than KS, the nonlocal calculation must be kept as cheap as possible. Ghasemi and K\"uhne attempted to learn the ``source" of the potential rather than the potential itself~\cite{Ghasemi_JCP2021}, which could alleviate the nonlocality.

A remarkable mention goes to a recent result by Zhang et al.~\cite{Zhang_NatComputSci2024}, in which they claim that the total energy difference of ML OF-DFT from KS-DFT was well under the chemical accuracy for broad unlearned molecules. They used the atom-centered density descriptor and developed a complicated NN architecture using the transformer~\cite{Turner_arxiv2023}, which is rapidly gaining popularity in various ML usages for its performance. This achievement may be the cornerstone on which the OF-DFT can next aim at true chemical accuracy.

\subsubsection{Exchange-correlation energy}
Fitting of the exchange-correlation energy by ML dates back to 1996 by Tozer, Ingamells and Handy~\cite{Tozer_JCP1996}. For small molecules composed of the first and second row elements, they calculated accurate charge densities using coupled clusters theory, solved the inversion problem of the KS equation~\cite{ZMP_PRA1994} to obtain the KS potential, fitted a NN model function, and applied the optimized functional to unseen molecules (still restricted to the second-row) to see its performance. They found that the optimized LDA functional deviates from an analytical LDA model to better fit the data. Thanks to the correct asymptotic behavior of the training KS potential, their NN LDA functional yields significant improvement for the ionization energy for the restricted group of test systems. Surprisingly, this study has already established the template design of the modern ML-DFT developments, as described later. 

Other researchers exploited Becke's three parameter hybrid form (Sec.~\ref{sec:semiempirical}) by making the mixing parameters $(a_{0}, a_{\rm x}, a_{\rm c})$ [Eq.~(\ref{eq:Becke_formula})] system-dependent. Zheng et al.~\cite{Zheng_CPL2004} implemented in 2004 a small NN which relates $n({\bf r})$ calculated with the B3LYP to $(a_{0}, a_{\rm x}, a_{\rm c})$ for correcting $E_{\rm xc}$ so that the reaction energies of the reference molecules are well reproduced, showing fair improvement in the test accuracy relative to B3LYP. Similar methods, where a small number of parameters entering the hybrid exchange-correlation functionals are derived from lower-level theories, have been further developed more recently~\cite{Liu_JPCA2017,Khan_2024arxiv}. This is not strictly DFT since the parameters are explicitly dependent on atomic numbers. These are modest usages of ML, whereas we showcase more drastic use below.

Exact fitting of the fully nonlocal $E_{\rm xc}[n]$ is interesting as it may help to infer the analytic properties of the exact exchange and correlation. For one-dimensional hydrogen chains, Li and coauthors fitted the exact exchange-correlation energy $E_{\rm xc}[n]$ calculated with the density matrix renormalization group~\cite{Li_Wagner_PRB2016} with KRR, showing that the test error rapidly converges to zero by increasing the training data. Nagai et al., and separately, Gavini et. al. ~\cite{Nagai_JCP2018, GaviniXC} performed a fitting of the fully nonlocal exchange-correlation potential $V_{\rm xc}(\mathbf{r})$ in an extremely simple one-dimensional model derived from the exact diagonalization and KS inversion, as a vector-to-vector mapping ${\bf n}\rightarrow {\bf V}_{\rm xc}=\{V_{\rm xc}(x_{1}),V_{\rm xc}(x_{2}), \cdots\}$ by the NN. Such exact fitting becomes demanding when going to three dimensions due to the difficulty of representing the density distribution in an efficient fashion. Progress in this problem is revisited below.

ML has also been helpful for development starting from the semilocal approximation, toward the chemical accuracy world but by going up paths forked from Jacob's Ladder (Fig.~\ref{fig:Ladder_boost}). The semilocal approximation is represented by the vector-to-scalar relation $\varepsilon_{\rm x, c}[n]({\bf r})\approx \varepsilon_{\rm x, c}({\bf g}({\bf r}))$ with e.g. ${\bf g}({\bf r})=(n({\bf r}),|\nabla n({\bf r})|)$ for the GGA. Extension of the descriptor vector ${\bf g}$ has been widely explored, exploiting the fact that the procedure for optimizing ML models is executable for any descriptor. 

A variety of nonlocal descriptors in the form Eq.~(\ref{eq:Nonlocal-n}) have been pursued, which are generalizations of the weighted density approximation for the exchange-correlation hole~\cite{Gunnarsson_PRB1976}. Schmidt et al.~\cite{Schmidt_JPCL2019} applied the convolutional NN form to the one dimensional model. Lei and Medford~\cite{Lei_Medford2019} introduced the decomposition of the three dimensional charge density by the Maxwell-Cartesian spherical harmonics and showed that the B3LYP exchange-correlation functional may be accurately fitted by a NN. Nagai et al.~\cite{Nagai_npjComputMater2020} implemented a NN model including a weighted density descriptor and applied to the KS equations, achieving accuracy comparable to a hybrid functional, without resorting to the generalized KS framework including the exact exchange operator (4th rung). Bystrom and Kozinsky~\cite{Bystrom_JCTC2022,Bystrom_PRB2024} constructed a convolution-type density descriptor that is invariant under the uniform scaling transformation $n({\bf r})\rightarrow \lambda^3 n(\lambda{\bf r})$, that is necessary for  modeling exchange functionals that guarantee exact scale invariance (see later).

The atomic decomposition of the charge density $n({\bf r})\approx \sum_{I:{\rm atom}}n_{I}({\bf r})$ has also been pursued for efficient descriptor design for the 3D charge density. The theory by Bart\'ok et al.~\cite{Bartok_PRB2013} on how to represent the atomic configuration surrounding a certain atom has been utilized by several researchers~\cite{Dick_JCP2019, Grisafi_ACS2019, Margraf2021}. They generally decompose the calculated $n(\mathbf{r})$ in an atom-centered basis and formulate rotationally covariant descriptors from the coefficients. A scheme of this kind allows for a representation of the entire distribution which is very sparse compared to those obtained from system-independent basis like plane-waves. The disadvantage is that one has to store the labels of atom types and positions: the functionals thus developed are in a strict sense not density functionals and may lose DFT's capability of generalizing across different types of atoms; yet they seem useful if the range of application is appropriately limited. ~\cite{Dick_NComm2020, Margraf2021}. Generalization to different atom types could be recovered by properly encoding the atomic position information, as shown in the case of OF-DFT~\cite{Zhang_NatComputSci2024}.

Not only the charge density, but \textit{any }quantities that are calculated using the converged KS Hamiltonian can also be taken as descriptors since they are all density functionals. The DeepMind21 (DM21) functional~\cite{Kirkpatrick_Sci2021} incorporates the local HF exchange energy density and LDA exchange energy density into the descriptors, the former of which is a functional of the KS orbitals ~\cite{PerdewPhysRev}. Riemelmoser et al.~\cite{Riemelmoser_JCTC2023} formulated an extremely nonlocal density descriptor calculable with the fast Fourier transform. Polak et al.~\cite{Polak2024_chemrxiv} adopt artificial densities calculated from unoccupied KS states as descriptors for the correlation energy density. A more data-based approach to the descriptor design would be to use the dimensionality reduction techniques as preprocessers of the functional input. Gong and coworkers~\cite{Gong_DigDis2023} trained an NN encoder of density distribution in three dimensions so that it can relate the distributions matched by the translation and scale transformation and utilized the encoded features as descriptors of the exchange energy.  

\begin{figure}[t]
\sidecaption
\includegraphics[scale=.35]{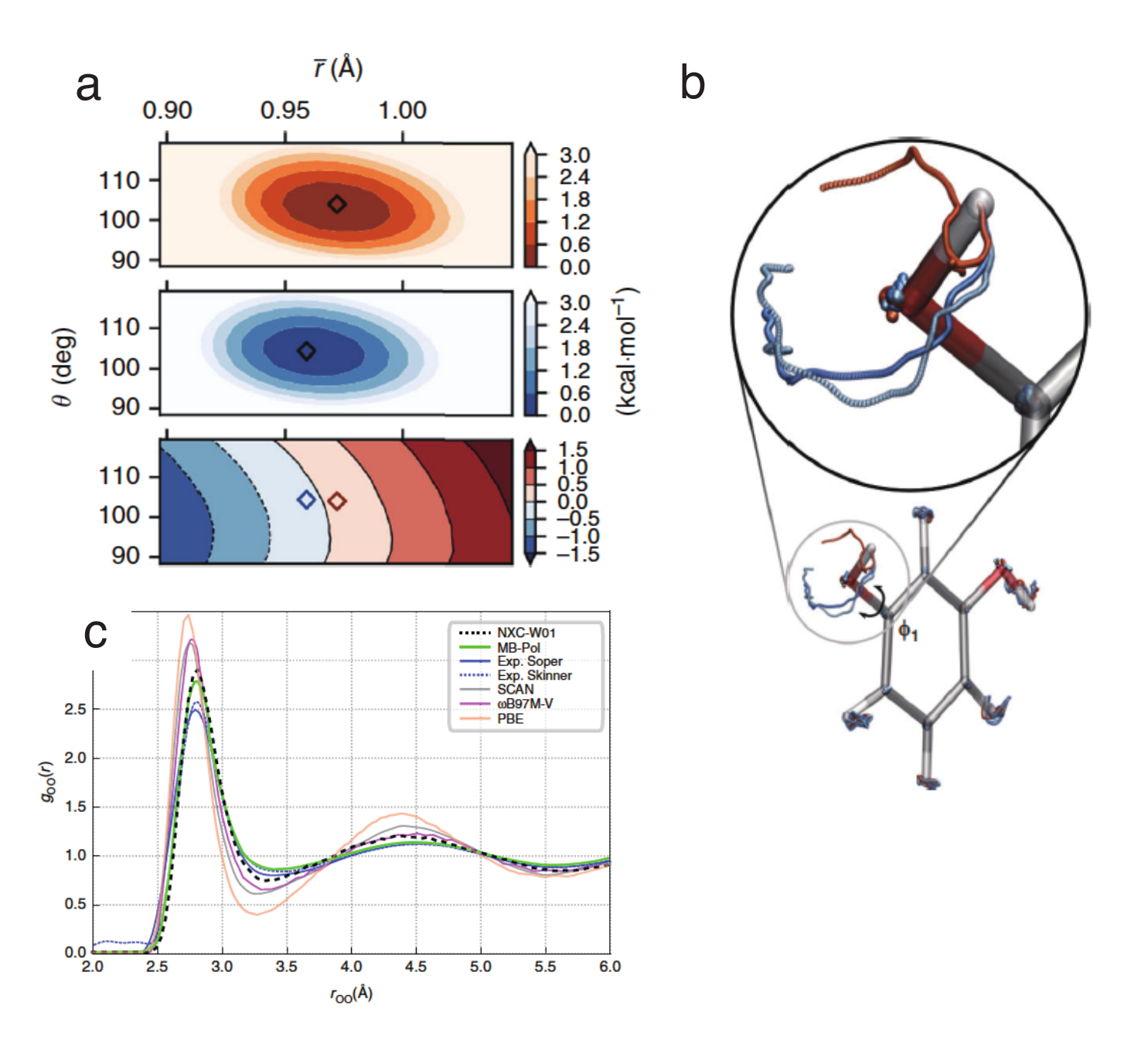}
\caption{Dynamics on accurate potential surfaces by $\Delta$-DFT. {\bf a} The potential for water molecule calculated by the baseline theory (GGA, top) is corrected to the surface of CC accuracy (middle). The correction ML term (bottom) shows smooth and modest behavior, which corrects the minimum from red to blue diamond. Taken from Ref.~\cite{Bogojeski2020} under the terms of the CC BY 4.0 license. {\bf b} The dynamics of resorcinol on the accurate surface generates rotational trajectories of a branch (light and dark blue), different from the GGA force field (red). Taken from Ref.~\cite{Bogojeski2020} under the terms of the CC BY 4.0 license. {\bf c} The radial distribution function for O-O in water from molecular dynamics, calculated by a ML nonlocal $E_{\rm xc}$ trained by the water trimer (``NXC-W01" in legend). Taken from Ref.~\cite{Dick_NComm2020} under the terms of the CC BY 4.0 license.}
\label{fig:CC_pot_surface}       
\end{figure}

\subsubsection{Baselining and $\Delta$-learning}
One key to succesful performances of ML-DFT is to leverage the wealth of intuition around human-made functional forms in concert with ML. One methodology for doing so is $\Delta$-learning; in which the ``baseline" of the model is an existing analytical functional form. Then, the \textit{difference} between the analytical form and the reference data (i.e the $\Delta$) is learned. Very often, these methods are presented in modest fashion (sometimes in the supplementary materials rather than the main text) despite their pivotal importance. We do not delve into the details, but discuss some general strategies.


In most of the ML applications based on the semilocal-to-nonlocal strategy, the model definition relies on the traditional decomposition or factorization of the exchange and correlation energies such as Eqs.~(\ref{eq:GGA-exchange}) and (\ref{eq:GGA-correlation}). These definitions are useful for applying constraints on the machines as detailed later.

Ramakrishnan et al.~\cite{Ramakrishnan2015} put forward a $\Delta$-ML concept, represented by the following formula
\begin{eqnarray}
    P_{\rm t}({\bf d}_{\rm t}) \approx P_{\rm b}({\bf d}_{\rm b})+\Delta P({\bf d}_{\rm b}).
\end{eqnarray}
Here, one wants to learn a target (``t") property $P_{\rm t}$ as a function of descriptors of target value ${\bf d}_{\rm t}$, assuming that the descriptor value is also determined self-consistently referring to $P_{\rm t}$ (for example $P$ and ${\bf d}$ may be the total energy and optimum atomic positions of molecules, respectively). In the DFT context ${\bf d}$ can be $n(\bf r)$~\cite{Sinitskiy2018_arxiv,Bogojeski2020}. For the learning task, one chooses a baseline theory ``b" that is calculated with lower computational costs and lets the machine learn the residual $\Delta P$ as a function of approximate value of descriptor ${\bf d}_{\rm b}$ derived from the baseline theory. This approach is straightforwardly applied to Jacob's ladder, to learn the higher level energies as a functional of $n(\mathbf{r})$ derived from lower-level theories. Thanks to this, the calculation of the total energy with  CCSD accuracy has been achieved with  DFT cost~\cite{Bogojeski2020}. The remaining error due to the incorrect ${\bf d}_{\rm b}$ is related to the density-driven error concept~\cite{Kim_PRL2013}, which we revisit later in this text.

\subsubsection{Derivatives of the functionals}
Similarly to the case of $T_{\rm s}[n]$, using the trained $F_{\rm model}[n]$ for solving the KS equation requires some care. Although in  $F_{\rm model}[n]$ we are able to take arbitrary descriptors, to use it for calculating $V_{\rm Hxc}({\bf r})=\delta F_{\rm model}[n]/\delta n({\bf r})$ the descriptors have to be formally differentiable, though a large class of descriptors written in the convolutional forms can be differentiated~\cite{Sahoo_ChemPhysChem2024}. Compared with the $T_{\rm s}[n]$ case, the effect of the numerical noise in $\delta F_{\rm model}[n]/\delta n({\bf r})$ may be mitigated by the explicit use of the kinetic energy operator $\hat{T}=-\nabla^2/2$~\cite{Nagai_JCP2018}. Still, stabilizing the KS cycles would be essential for reliable applications. Regularizations  based on perturbation theory~\cite{Kirkpatrick_Sci2021} and taking unconverged KS steps into the loss~\cite{Li_PRL2021} have been proposed.

Another important derivative is the force on atoms ${\bf F}_{\rm I}=-\partial E_{\rm GS}[n]/\partial {\bf R}_{\rm I}$, which is essential for performing a molecular dynamics simulation. If one can learn an accurate charge density, for the Coulombic system, the Hellmann-Feynman force formula ~\cite{Hellmann_Feynman_PRB1939}
\begin{eqnarray}
  {\bf F}_{\rm I}=-\int d{\bf r}n({\bf r})\frac{Z_{I}e^2}{|{\bf R}_{I}-{\bf r}|^3}\left({\bf R}_{I}-{\bf r}\right)+\sum_{I'\neq I} \frac{Z_{I}Z_{I'}e^2}{|{\bf R}_{I}-{\bf R}_{I'}|^3 }\left({\bf R}_{I}-{\bf R}_{I'}\right)
\end{eqnarray}
enables us to calculate the force with the same accuracy. Dick and Fernandez-Serra~\cite{Dick_NComm2020} thus executed molecular dynamics with an accurate force field for liquid phases of small molecules, obtaining accurate radial distribution functions. For frameworks that do not target accurate charge density, some additional precalculation is required such as the atomic-position derivative of the atom-centered basis set~\cite{Chen_JCTC2021} similar to the Pulay term~\cite{Pulay_MolPhys1969}. Bogojeski et al. demonstrated accurate MD simulation of a molecule, by intermittently correcting the baseline force by the correction term calculated with the finite-difference method to the ML higher-rung energy~\cite{Bogojeski2020}.

\subsubsection{Other quantities}
The quantity of central interest in DFT calculations is the energy, while other quantities at the ground state, including responses functions, are also functionals of $n(\mathbf{r})$. Moreno, Carleo and Georges~\cite{Moreno_PRL2020} attempted the fitting of the many body wavefunction and two-body correlation function as functionals of $n(\mathbf{r})$, implying that those high-dimensional quantities are on a reduced latent dimension of $n(\mathbf{r})$ in line with the Hohenberg-Kohn theorems. Applications to single-particle excitations from the ground states~\cite{Bystrom_JCTC2024} and real-time dynamics~\cite{Suzuki_PRA2020} have also been demonstrated.

\begin{figure}[t]
\sidecaption
\includegraphics[scale=.45]{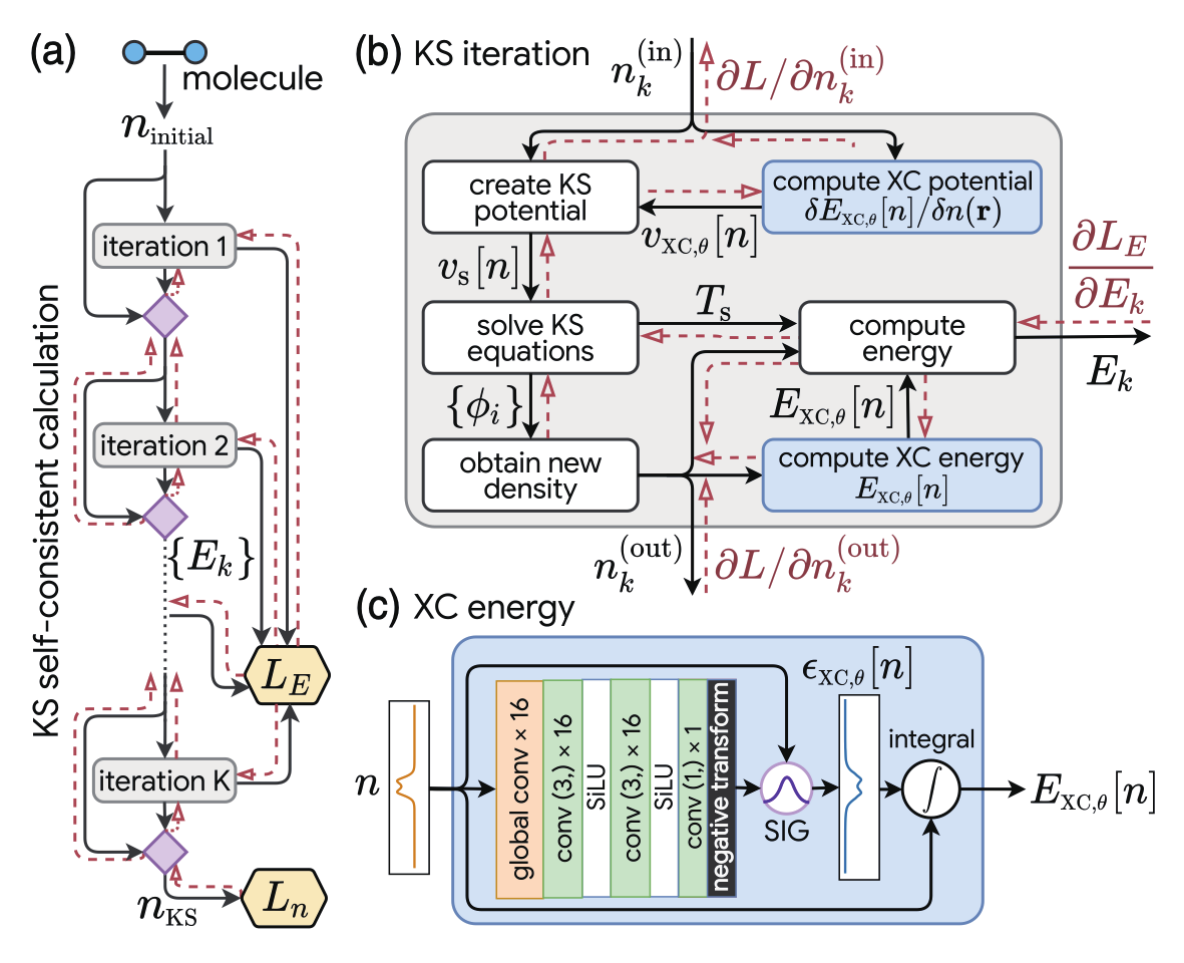}
\caption{{\bf (a)--(c)} Learning through the differentiable KS equation, taken from Ref.~\cite{Li_PRL2021} under the terms of the CC BY 4.0 license. The exchange-correlation energy [{\bf (c)}] is modeled. The loss function is defined with the energies at unconverged steps and converged charge density [{\bf (a)}]. The parameter derivatives backpropagate through the iterative calculations by the chain rule [{\bf (b)}]. This helps produce a functional which converges.}
\label{fig:differential_KS}       
\end{figure}

\subsection{Learning through the KS cycle}
For the training of the universal functionals, one may desire to refer to quantities other than the functional itself, such as $n(\bf r)$, from the expectation that the condition number in the model optimization should be increased and agreement with the observable quantities are improved. This is also a matter of curiosity that whether people can improve accuracy of $n(\mathbf{r})$, the fundamental variable of DFT, even if accurate observation of it is difficult. The task appears tricky when the quantities are not dependent explicitly on the model parameters but dependent only through the KS solution using parametrized functionals. Nagai et al.~\cite{Nagai_npjComputMater2020} trained the exchange-correlation energy density with the loss function defined by $n(\mathbf{r})$. They optimized the parameters by derivative-free Monte Carlo sampling and indeed showed that the training on only three molecules can generalize to 147 molecules, although the optimization was extremely time-consuming. Similar generalization from a few to hundreds of molecules has been demonstrated by Wang et al.~\cite{Wang_JCP2022}, showing the efficiency of the training on $n(\mathbf{r})$. Thankfully, since then, the usual gradient-based optimization has been already made applicable as we sketch below. 

Chen et al.~\cite{Chen_JCTC2021} took another approach which may be applied to learning of other quantities calculated through the self-consistent KS solution. Let $Q$ be any quantity to learn which is in theory formulated with the outputs of the converged KS equation; e.g., $Q=n$. Their learning scheme is then expressed as the following minimization\footnote{We omit the force term in the original work as it is not necessary for the present discussion.}
\begin{eqnarray}
    \min_{\omega}\sum_{i=1}^{N_{\rm data}}\left[E_{\rm data}^{(i)}-E_{\rm KS}^{(i)}(\omega)\right]
    ,
\end{eqnarray}
where $E_{\rm KS}^{(i)}$ denotes the ground state energy from {\it a modified} KS equation modeled with a set of parameters $\omega$:
\begin{eqnarray}
    E_{\rm KS}^{(i)}={\rm min}_{\{\varphi_{j}\},\sum_{j:{\rm occ.}}|\varphi_{j}|^2=n({\bf r})}\left[F_{\omega}[n]+E_{\rm ext}^{(i)}[n]+\lambda_{Q}D[Q[n],Q_{\rm data}^{(i)}]\right]
    .
\end{eqnarray}
The parametrized universal function $F_{\omega}$ is assumed to be the target to model. Here $D[Q,Q']$ is a given nonnegative function which becomes zero only when $Q$ and $Q'$ agree., whereas $\lambda_{Q}$ is the regularization parameter. As a key trick, they switched the order of the minimizations from ${\rm min}_{\omega}\left({\rm min}_{\{\varphi_{j}\},\sum_{j:{\rm occ.}} }\cdots \right)$ to ${\rm min}_{\{\varphi_{j}\},\sum_{j:{\rm occ.}}} \left( {\rm min}_{\omega}\cdots \right)$. The gradient optimization of the parameter $\omega$ is then executed as usual with a fixed set of orbitals $\{\varphi_{j}^{(i)}\}$ where $i$ and $j$ run over the training systems and occupied states, respectively. The orbitals are infrequently updated by the parametrized KS equation including a penalty potential $V^{(i)}_{\rm Q}=\lambda_{Q}\delta D[Q[n],Q^{(i)}_{\rm data}]/\delta n$. Those optimizations are repeated until convergence, after which the optimized $F_{\omega}$ is in hand that is tuned to best yield $Q$ as well as $E$ for given training systems.

The idea that the KS self-consistent calculation can be made differentiable is remarkable~\cite{Li_PRL2021}. Any mathematical algorithm implemented in a programming language is  executed by sequences of fundamental operations; addition, subtraction, multiplication, and division. From this, even complicated algorithms can be described in terms of composite functions formed by those fundamental operations and therefore its derivatives can be formulated by the chain rule. The differentiable implementation of the KS self-consistent cycle has then emerged and enabled us to differentiate the physical quantities calculated via the self-consistent KS solution. For a one dimensional model, Li et al.~\cite{Li_PRL2021} optimized the NN-model exchange-correlation energy with respect to the loss function referring to the converged charge density. Soon after Kasim and Vinko~\cite{Kasim_PRL2021} and Dick and Fernandez-Serra~\cite{Dick_PRB2021} demonstrated a three-dimensional implementation for the optimization of the NN-model meta GGA. This technology is widely applicable to ML-DFT as it extends the variety of the learning framework, though the differentiable programming has its own challenges as it requires large memory to store the numerical derivatives.

\begin{table}[ht]
  \centering
  \caption{Performance of the ML meta GGA model (``pcNN-based") with exact constraints; data taken from Ref.~\cite{Nagai_PRR2022}. The model was trained on three molecules. ``NN-based" represents the model from Ref.~\cite{Nagai_npjComputMater2020}, where the constraints were not imposed. (a) Atomization energy benckmark on 144 untrained molecules. (b) Optimum lattice constants of 48 solids. Parentheses in column ``NN-based" indicates that 6 systems were excluded because of ill convergence of the KS cycle. (MAE, mean absolute error; ME, mean error; SD,  standard deviation of signed error)}
  \label{tab:PCNN_test}
  \begin{subtable}[t]{0.45\textwidth}
    \centering
    \caption{Atomization energies of molecules}
  \begin{tabular}{lccc}
    \toprule
     & \textbf{MAE} & \textbf{ME} & \textbf{SD } \\
    \textbf{XC} &(kcal/mol)&(kcal/mol)& (kcal/mol) \\
    \midrule
    PBE        & 17.3 & 16.2 & 13.1 \\
    SCAN       &  6.2 & -4.5 &  5.7 \\
    NN-based   &  4.8 &  1.8 &  6.3 \\
    pcNN-based &  3.6 &  0.3 &  4.5 \\
    \bottomrule
  \end{tabular}
  \end{subtable}
  \hfill
  \begin{subtable}[t]{0.45\textwidth}
    \centering
    \caption{Optimized lattice constants of solids}
    \begin{tabular}{l@{\quad}c@{\quad}c@{\quad}c}
    \toprule
     & \textbf{MAE } & \textbf{ME} & \textbf{SD} \\
    \textbf{XC} &  (m\AA)  & (m\AA)& (m\AA) \\
    \midrule
    PBE        & 38.1  & 33.9  & 44.2  \\
    SCAN       & 22.3  & -7.5  & 28.5  \\
    NN-based   & (22.9) & (0.8) & (32.0)\\
    pcNN-based & 19.1  & -2.5  & 26.5  \\
    \bottomrule
  \end{tabular}
  \end{subtable}
\end{table}


\subsection{Machine-learning and exact conditions}
Even with the recent flourish of ML-DFT, the traditional analytical approaches remain important. Rigorous equalities and inequalities from the latter are useful for regularizing the behavior of the trained ML models when applied to less familiar systems. Usage of those as constraints have been explored as summarized below.

Uniform scaling equalities and inequalities are properties satisfied by the exact functional~\cite{Levy_Perdew_scaling_PRA1985}. Under uniform scale transformation, the charge density transforms as:
\begin{eqnarray}
    n({\bf r})\rightarrow n_{\rm \gamma}({\bf r})=\gamma^D n(\gamma {\bf r}),
\end{eqnarray}
where $D$ denotes the spatial dimension. With this transformation, the exact kinetic energy transforms as $T_{\rm s}[n]\rightarrow T_{\rm s}[n_{\rm \gamma}]=\gamma^2 T_{\rm s}[n]$ and the exact exchange obeys $E_{\rm x}[n]\rightarrow E_{\rm x}[n_{\gamma}]=\gamma E_{\rm x}[n]$. Hollingsworth et al.~\cite{Hollingsworth_JCP2018} proposed a preprocess to transform the input $n(\bf r)$ to a definite scale $n(\mathbf{r})\rightarrow \tilde{n}(\mathbf{r})$ so that the density distributions related by the scaling transformation are identified. The machine then learns only the processed $\tilde{n} 
(\bf r)$ and the actual energy is calculated via the exact scaling transformation at the postprocess. A loosened version of the preprocess has also been proposed by Gong and coworkers~\cite{Gong_DigDis2023}. They trained an NN encoder of density distribution in three dimensions so that it can relate the distributions matched by the translation and scale transformation and utilized the encoded features as descriptors of the exchange energy. Bystrom and Kozinsky formulated a weighted-density type descriptor that is explicitly invariant with respect to the scale transformation~\cite{Bystrom_JCTC2022,Bystrom_PRB2024}. Using this, they implemented an $E_{\rm x}$ that is forced to satisfy the scaling invariance, and is succifiently accurate to replace exact exchange in global hybrids ~\cite{Polak2024_chemrxiv}.
\\

Various asymptotic formulas and inequalities are available~\cite{Sun_PRL2015} as constraints on an ML model. Building models which strictly satisfy such constraints--and therefore avoid overfitting to training data--is now possible ~\cite{GGAPBE1996}. Cuierrier {\it et al.}~\cite{Cuierrier_JCP2021} and Sparrow {\it et al.}~\cite{Sparrow_JPCL2022} included the variance from the exact constraints to the loss function, with which the model is regularized to comply with those approximately. Lagrange interpolation to the end point of an ML model~\cite{Nagai_PRR2022} can force the model more strictly to converge to the desired asymptotic limits. Inequality constraints can also be enforced by using any bounded activation functions at the output~\cite{Dick_PRB2021}. Nagai et al.~\cite{Nagai_PRR2022} have shown that, by designing an ML model which recovers the uniform electron limit and other constraints used in the strictly constrained and appropriately normed (SCAN) meta GGA~\cite{Sun_PRL2015}, the resulting functional is reliably applicable to solid systems, even when it was trained only on molecules (Table~\ref{tab:PCNN_test}).

Linear dependence of the total energy on total number/spin of electrons is a significant exact constraint.  DFT applied to non-integer $N=\int d{\bf r} n({\bf r})$ by defining the non-integer ground state as a linear superposition of the ground state density matrices with integer electrons~\cite{Janak_PRB1978, Perdew_PRL1982}. Ultimately, this is simply the $T \to 0$ limit of Mermin's finite-temperature generalization of the Hohenberg-Kohn theorems to the grand canonical ensemble ~\cite{Mermin65Temp}. From this definition, the ground state energy as a function of $N$, $E_{\rm GS}(N)$, is linear between integer values of  $N$~\cite{Perdew_PRL1982,Perdew_PRL1983}. This linearity condition, which is usually violated by approximate functionals, is found to be tied to a proper description of charge transfer within the system with definite $N$, affecting the accuracy of molecular dissociation and the charge gap~\cite{Cohen_Sci2008,Cohen_JCP2008}. Researchers implemented the charge-spin linearity conditions to the model by construction ~\cite{Gedeon_MLST2021} or training of the model with data enforcing the linearity~\cite{Kirkpatrick_Sci2021}. In particular, the DM21 functional by Kirkpatrick and colleagues~\cite{Kirkpatrick_Sci2021} succeeds in reproducing the dissociation curves of various molecules, passing the stringent test for DFT. Unfortunately, that approximation is costly to evaluate, and has significant convergence issues.

An important insight on using the exact constraints can be found in Hollingsworth et al.~\cite{Hollingsworth_JCP2018}. In their test of learning $T_{\rm s}[n]$ compliant to the scaling relationship in one dimensional models, significant improvement was seen in the case of the Hooke's atom, whereas for the stretched 1D H$_2$ the performance improvement was marginal. This is due to the failure of learning the size consistency because of the scaling preprocess $n(\mathbf{r})\rightarrow \tilde{n}(\mathbf{r})$. The effective use of the constraints thus requires knowledge of which constraints regularize which physics~\cite{Pederson_JCP2023}. It would be of great interest to apply such conditions with more modern ML methods to quantities such as the exact exchange. 

\section{Where do we go from here?}
There has been much progress in the proofs of new concepts in the DFT thanks to ML. In this section we attempt to summarize the achievements and raise an issue of the importance for the DFT framework deduced from the ML applications.

\begin{figure}[t]
\sidecaption
\includegraphics[scale=.20]{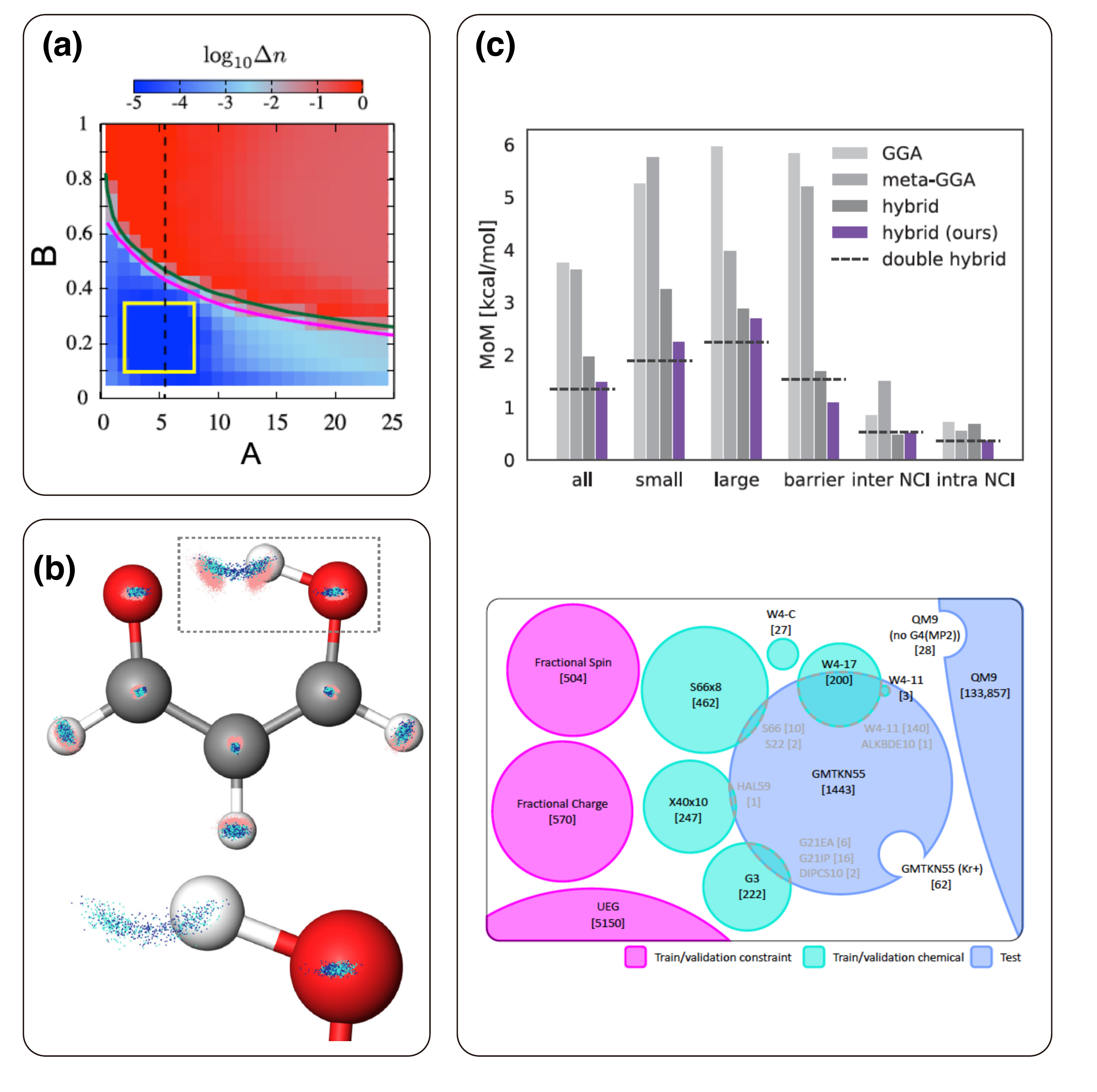}
\caption{Suggestive results concerning generalization. {\bf a} Clear example of failed generalization. In a 1D model the ionic potential is controlled by two parameters $A$ and $B$. The machine exclusively trained on data in a single phase (yellow frame) does not generalize beyond the phase boundary. Reprinted from R. Nagai, R. Akashi, S. Sasaki and S. Tsuneyuki ``Neural-network Kohn-Sham exchange-correlation potential and its out-of-training transferability", J. Chem. Phys. \textbf{148}, 241737 (2018) with the permission of AIP Publishing~\cite{Nagai_JCP2018}. {\bf b} Configurations of malonaldehyde. The model trained with red points generate geometries including the proton transfer when applied to MD (blue), that agree well with those generated by the first-principles MD. Taken from Ref.~\cite{Brockherde_NatComm2017} under the terms of the Creative Commons CC BY. {\bf c} (top) Performance of the DM21 functional reaching the accuracy of double hybrid, at the cost of a single hybrid. (bottom) The training and test data, whose overlap seems small. From J. Kirkpatrick \textit{et al.}, ``Pushing the frontiers of density functionals by solving the fractional electron problem", Science \textbf{374}, 1385 (2021) ~\cite{Kirkpatrick_Sci2021}. Reprinted with permission from AAAS.}
\label{fig:generalize}       
\end{figure}

\subsection{Generalization: Summary}
\label{subsec:generalization}
We classify the types of generalization presented in published papers:   \textit{Configurational generalization}~\cite{Snyder_PRL2012,Snyder_JCP2013,Snyder_IJQC2015,Li_JQC2016,Li_Wagner_PRB2016,Hollingsworth_JCP2018,Nagai_JCP2018,Brockherde_NatComm2017,Bogojeski2020,Dick_NComm2020,Li_PRL2021,Schmidt_JPCL2019,Gedeon_MLST2021,Moreno_PRL2020,Margraf2021,Denner_PRR2020,Ryabov_SciRep2020,Seino_CPL2019,Fujinami_CPL2020,Meyer_JCTC2020,Ryczko_JCTC2022,Golub_PCCP2019,Manzhos_JCP2020,Manzhos_JCP2023,Remme_JCP2023,Mazo-Sevillano_JCP2023,Ghasemi_JCP2021} means generalization within a fixed system, where the training is performed for various atomic configurations and test is done for untrained configurations of the same system. \textit{Compositional generalization}~\cite{Tozer_JCP1996,Nagai_npjComputMater2020,Tsubaki2020,Dick_PRB2021,Kasim_PRL2021,Kirkpatrick_Sci2021,Nagai_PRR2022,Cuierrier_JCP2022,Riemelmoser_JCTC2023,Bystrom_PRB2024,Bystrom_JCTC2024,Pokharel_JCP2022,Casares_JCP2024,Mendonca_JCTC2023,Ryabov_SciRep2022,Kanungo_arxiv2024,Chen_JCTC2021,Chen_Yang_JCP2024,Kalita_JPCL2022,Seino_JCP2018,Imoto_PRR2021,Lueder_EleStr2024,Zhang_NatComputSci2024,Sun_Chen_PRB2024} denotes the case where the training is performed for a certain group of compounds and the test is done for other compounds not seen during training. The latter generalization is the very goal of DFT development, whereas the former generalization is also useful for thermodynamic methods like molecular dynamics.
Some studies indicate that ML can achieve configurational generalization by skipping rungs in Jacob's ladder while retaining  coupled cluster accuracy, based on the $\Delta$-learning framework. Bogojeski et al.~\cite{Bogojeski2020} implemented a KRR model that relates ${\bf n}$ to the energy difference between CC and GGA, $\Delta E_{\rm model}^{\rm CC-GGA}[n_{\rm model}^{\rm DFT}]$\footnote{The model accepts approximate $n_{\rm model}^{\rm DFT}$, which is calculated by another KRR model that relates the ionic potential and $n$ calculated from the semilocal DFT; the calculation of this term only can therefore be done by further reduced cost.}, and built models optimized for molecular H$_{2}$O, ethanol and resorcinol; all of which yielded accurate potential surfaces with errors $<$ 1~kcal/mol from the CC results. Dick and Fernandez-Serra~\cite{Dick_NComm2020} also built a model for the energy difference from the baseline functional (GGA) that is also which yields more accurate $n$  than the baseline approximation. Their model trained on the H$_2$O trimer was used for molecular dynamics with supercell containing 96 H$_{2}$O molecules, yielding very accurate radial distribution functions. Both studies attain chemical accuracy at GGA cost, skipping three steps of the ladder Fig.~\ref{fig:Ladder_boost}.

We can find some unexpected generalizations from the configurational studies, where the machines retain accuracy even in the regimes to which we would not anticipate extrapolation. In Brockherde et al.~\cite{Brockherde_NatComm2017}, a model was trained on malonaldehyde, where configurations on the verge of the proton transfer were not present. Nevertheless, it accurately described the proton transfer process when applied to the molecular dynamics (Fig.~\ref{fig:generalize}b), though less accurately than for other configurations. Generalization of this kind is what physics-oriented researchers hope to achieve in  ML applications, i.e. that the machine finds some apparent extrapolation is actually an interpolation, according to the latent space embedded in the training model. The configurational studies also give us clear insights where the models do not generalize and why. Nagai et al.~\cite{Nagai_JCP2018} examined a simple one-dimensional Hamiltonian with variable number of electrons bound in a potential. They showed that the machine trained solely on potentials which bind only one electron never generalizes to potentials which bind two (Fig.~\ref{fig:generalize}a), and vice versa. Closer analysis of such successful and failed generalizations in the literature may give us insights that are brought back to the analytical DFT development.

The compositional kind of generalizations are usually shown by the study design where researchers take training data for compound databases and divide them into training, validation and test groups. In such studies the DM21 functional~\cite{Kirkpatrick_Sci2021} should be worth highlighting (Fig.~\ref{fig:generalize}c). They prepared exchange-correlation energies with CCSD(T) accuracy for a training set consisting of 1161 molecular reactions and 1074 cases of fractionally charged or spin-polarized atoms from H to Ar. The exchange-correlation energy density was modeled as a functional of the semilocal density descriptor calculated with the B3LYP functional (fourth rung). Training was done with an additional regularization term in the loss that is intended for stabilizing the KS self-consistent loop. The trained NN model for the exchange-correlation energy density was tested by the KS self-consistent calculations for over 10$^5$ systems from the GMTKN55 and QM9 datasets, showing accuracy of double hybrid (fifth rung); one level skipped. The similarity between the training and test data seems low, which may indicate that the machine learned some latent dimensions in which the test systems are regarded interpolation. 

The most interesting goal for the compositional models would be to use it for exploration of new compounds. For this, the models have to attain some confidence when they are applied to undiscovered materials in \textit{in silico} , but this goal still seems distant. For instance, some ``out-of-paper" tests for DM21 reported untested systems where the DM21's accuracy declines, such as water clusters~\cite{Palos_JCP2022} and transition metal compounds~\cite{Zhao_PCCP2024}. Analyses of the latent space in the trained compositional functionals in some way would help human to judge if the functionals is applicable to any given undiscovered systems. As an related effort let us recall Pokharel et al.~\cite{Pokharel_JCP2022} which examine how broadly a functional which learned a few selected ``norm" systems, to which the model is assured to agree, may generalize. A more data-based approach to formulate the similarity measure between the databases is also interesting~\cite{Gould_ChemSci2024}.

\clearpage

\subsection{How to measure the accuracy of the density}
Ground-state DFT is a theory based on the ground-state charge density.  However, it really only uses the density as an auxillary variable.  Its chief use is to produce approximate energies for various different configurations of nuclear potentials.   That is what the thousands of applications that are published each year mostly do.

However, whether we need accurate densities for accurate machine learning can also be in question ~\cite{Medvedev_Science2017,Kim_PRL2013, Sim_JPCL2018, PaolaVMC}. As a matter of fact, there is much fascination about development of models that bypass DFT altogether, by formulating the machines with descriptors derived from atomic positions and compositions of the system concerned, from which it is implied that for well-controlled interpolation tasks we do not have to refer to the charge density. This is used to make both machine-learned potentials and also models for other properties, such as dipole moments, that we would usually extract
from integrals over the density.

\begin{figure}[t]
\includegraphics[scale=.45]{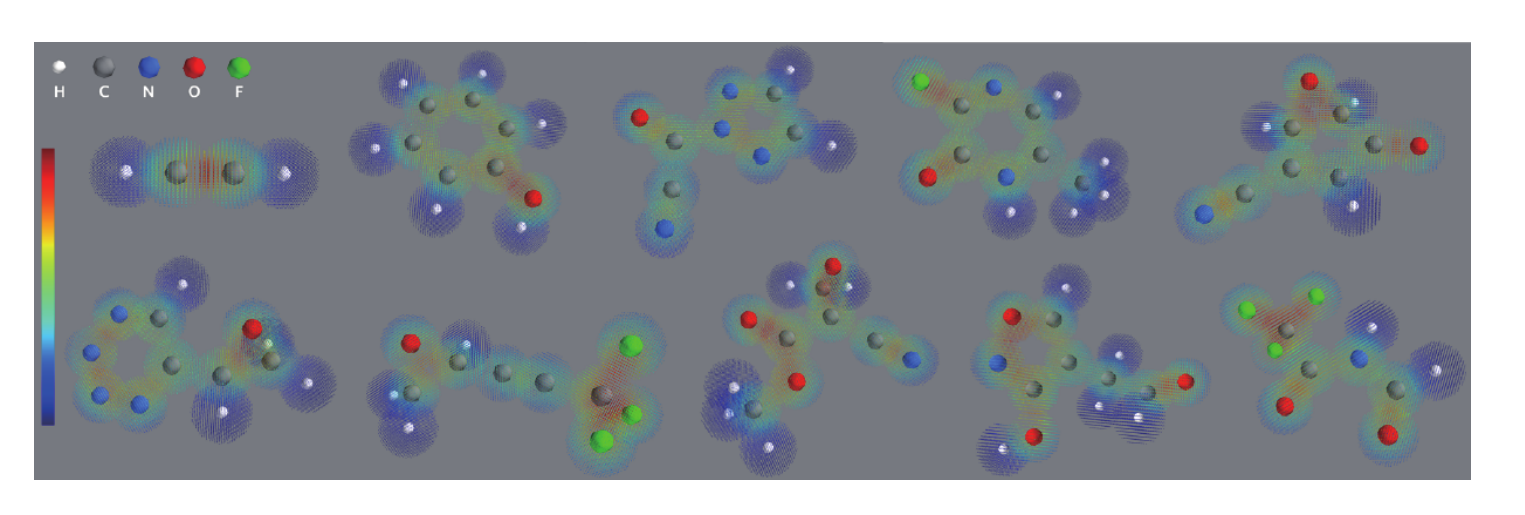}
\caption{``Charge density" as an intermediate field value in a trained machine, which was not directly referenced in the training, taken from Ref.~\cite{Tsubaki2020} under the terms of the CC BY 4.0 license.}
\label{fig:density_approx}       
\end{figure}

\begin{table}[ht]
  \centering
  \caption{Relevant dimension estimation of approximate densities for describing energies; data taken from Ref.~\cite{Bogojeski2020}.}
  \label{tab:density_relevant_dim}
  \begin{tabular}{c@{\quad}c@{\quad}c@{\quad}c@{\quad}c@{\quad}c}
    \toprule
    \textbf{Density/Energy} 
      & $E^\mathrm{DFT}_\mathrm{SML}$ 
      & $E^\mathrm{CC}_\mathrm{SML}$ 
      & $E^\mathrm{CC}_{\mathrm{SA\mathchar`-DFT}}$
      & $E^\mathrm{SAD}_{\mathrm{SA}}$ 
      & $E^\mathrm{CC}_{\mathrm{SA\mathchar`-SAD}}$ \\
    \midrule
    $n^\mathrm{DFT}$
      & 1196.56 & 1208.05 & 4335.93 & n/d    & n/d \\
    $n_s^\mathrm{DFT}$
      & 1162.68 & 1161.95 & 4010.18 & n/d    & n/d \\
    $n^\mathrm{SAD}$
      & 661.39  & 623.38  & 3639.97 & 677.66 & 374.98 \\
    \bottomrule
  \end{tabular}
  \vspace{1ex} 
  \\
  \textrm{n/d = not determined}
\end{table}

If we do not need too accurate $n(\bf r)$ for practical accuracy, it should further accelerate our development since the accurate calculation of $n(\bf r)$ beyond Hartree-Fock or semilocal DFT for training is itself a demanding task, and also the storage of accurate $n(\bf r)$ requires much more disk resources than energy-related observables.

On this issue let us refer to Brockherde et al.~\cite{Brockherde_NatComm2017}. Two machines were tested there: one is a machine that maps the one-body potential $V_{\rm ext}$ to the total energy (termed machine-learning Kohn-Sham map), and the other is a composite machine in which two machines respectively map $V_{\rm ext} (\bf r)$ to $n(\bf r)$ (termed machine-learning Hohenberg-Kohn map) and $n (\bf r)$ to $E$. The latter $n(\mathbf{r })\rightarrow E$ machine then learns the functional relating the approximate $n(\bf r)$ to the total energy, say, $E[n_{\rm approx.}]$, that is of course different from the exact density functional. Interestingly, the latter machine $V_{\rm ext}(\mathbf{r}) \rightarrow n(\mathbf{r}) \rightarrow E$ achieved reduction of the test errors more easily. This scheme has been further sophisticated as $\Delta$-DFT for learning the exchange-correlation energy with the coupled-cluster accuracy~\cite{Bogojeski2020}.

Tsubaki and Mizoguchi~\cite{Tsubaki2020} also showed an interesting demonstration. They developed a machine composed of two functionals that commonly use the LCAO coefficients of the orbitals as input. One relates the coefficients to the atomization energy as the NN output. The other first calculate the charge density $n(\mathbf{r})$ by the exact KS formula (Eq.~\ref{eq:KS-eq}) and next calculate the external potential from the calculated $n(\mathbf{r})$ through a NN. The two NNs are optimized so that the end data, atomization energy and the potential,  agree well with the training data. On training it does not appear to matter whether the intermediate $n(\mathbf{r})$ is accurate; however, the trained machine gave plausible (but of course inaccurate) $n(\mathbf{r})$ (Fig.~\ref{fig:density_approx}). Incorporating the orbital-to-$n(\mathbf{r})$ relation into the machine thus seems effective for efficient learning of the energy, even if the machine does not learn $n(\mathbf{r})$ directly. 

Bogojeski {\it et al.}~\cite{Bogojeski2020} examined the accuracy of $E_{\rm xc}[n]$ with different levels of approximation to $n(\mathbf{r})$. They applied the relevant dimension estimation method in the data science~\cite{Braun_JMLR2008} and calculated the signal-noise ratio, finding that, as $n(\mathbf{r})$ becomes inaccurate, the noise in the input-output pair becomes larger (Table~\ref{tab:density_relevant_dim}), which indicates that an accurate $n(\mathbf{r})$ carries more information relevant to $E_{\rm xc}$.

For understanding the issue of learning accurate densities, the framework of density-corrected DFT (DC-DFT) ~\cite{Kim_PRL2013} should prove useful. A general analysis (formalized in a broader context than ML alone) allows for the decomposition of the error of a DFT calculation into two pieces: (i) the error due to the approximation used for the form of $E_{\rm XC}$ and (ii) the error due to the approximate density which is found by minimizing the approximate functional. Letting $\tilde{E}$ and $\tilde{n}(\mathbf{r})$ denote the approximate energy functional and density respectively; with $E$ and $n(\mathbf{r})$ being the exact functional and density, the total error can be written as:

\begin{eqnarray}
    \Delta E = \tilde{E}[\tilde{n}]-E[n] = (\tilde{E}[n]-E[n]) + (\tilde{E}[\tilde{n}]-\tilde{E}[n])
\end{eqnarray}
where $\tilde{E}[n]-E[n]$ is called the ``functional driven error" and $\tilde{E}[\tilde{n}]-\tilde{E}[n]$ is called the ``density driven error". In the majority of cases, the functional-driven error dominates, but in some important systems, such as those containing radicals, the density-driven error is dominant. A DC-DFT calculation specifically seeks to reduce the density-driven error by evaluating an approximate functional on a \textit{different} density than its variational minimum. Such a calculation is refered to as ``inconsistent" in the literature, as opposed to the \textit{self}-consistent solution of KS equations ~\cite{WassermanInconsistent}. Often, Hartree Fock densities are used for this purpose (such a calculation is called HF-DFT), and for many molecules, this greatly reduces the density-driven error~\cite{Sim_JPCL2018}. While HF-DFT has so far been the main practical manifestation of density correction, the principle of DC-DFT is much broader, and is not restricted to Hartree Fock densities alone. This opens a path for constructing highly accurate ML functionals by learning the energy as a functional of a non-self-consistent density.

\section{The future?}

The application of machine learning in the physical sciences is producing a paradigm shift in these fields, creating many new capabilities that are driven by data and compute power.  This was recognized by the Nobel prize in both physics and chemistry in 2024.  A generation from now, ML will be as much a part of science as calculus is today.

In the specific area of electronic structure, the greatest impact so far has been the development of meachine-learned potentials (MLPs), which are already being used to generate new science (see other chapters).   A great example is the MD simulation of a shock wave passing through a slab of carbon, with 18 billion carbon atoms~\cite{Nguyen-Cong_SC2021}. Such a calculation is both impossible with DFT and not desirable, if (essentially) the exact same results are produced by the MLP.  More recently, foundational models, trained on the entire periodic table, have appeared~\cite{Batatia_arxiv2024}. While less accurate and robust than is desirable, the ability to perform much larger simulations on almost any phase of matter will totally alter all sciences it is applied to, at both the computational and experimental levels.  Such codes will be used by theorists and experimentalists alike.

Where does this leave DFT itself?   Will it be no more than the starting point for such models, by generating their training data? The answer is yes and no.  It will be used to generate training data, but also used to validate the models in places where they fail.   It also seems likely that it will also be used to fine-tune to improve accuracy in specific situations.   This is very analogous to how high-level quantum chemical and sometimes Monte Carlo data are used to benchmark and check DFT calculations today.

Which brings us to the topic of this chapter.  The use of ML to directly improve approximate density functionals.  We have discussed two specific functionals, the KS kinetic energy functional and the XC functional.  While important pioneering steps have been made so far, no all purpose approximation has yet appeared that has been widely adopted by the user community.  The hopes of DM21 have been dashed by its high computational cost and the difficulties with convergence.

We also want to make another distinction.  Some work is aimed at taking a human-designed approximate form, and trying to improve over human attempts with that approximate form.   Examples are GGA's and meta-GGA's and hybrids.   An early example is the $\omega$B97 series of functionals~\cite{Chai_JCP2008,Mardirossian_MolPhys2017}, which were data driven, and essentially regularized fits to avoid overfitting.  Such attempts are unlikely to work on problems where the human-designed counterparts fail, e.g. for strongly correlated systems.   What they aim to do is improve accuracy on the systems that already work.  A clever aspect of DM21 was to take a (relatively) little-explored human-designed form, so both improved accuracy and new capabilities could both be possible.

On the other hand, the rest of the work creates functionals that look nothing like anything suggested by people, i.e., which use density inputs that are not the usual semilocal suspects.   Some even use the entire density itself, allowing the possibility of creating the exact functional defined by HK long ago.   With enough data in the right form, it seems one can drive down errors far below the current forms, and create functionals that work even for strongly correlated systems.  

So why has this approach not been more broadly explored?   Ultimately, it creates a new and unexpected problem:   it tends to generalize poorly. The amount of training data needed to allow generalization across chemical species appears daunting and impractical.   This difficulty can be traced back to use of the {\em entire} density.  A tiny change in a density anywhere can, in principle, substantially change the ML energy.   In practice, this almost never happens, but occasionally it is important.  It is very difficult for a general scheme to reproduce this behavior.  It is useful to muse on the generality of the KS scheme with the simple LDA approximation, using only the density at a single point to generate the XC energy-- it has zero dependence on what the chemical or material species is.

This suggests that for such methods to become more useful, and go beyond proofs-of-principle, one needs to identify a limited set of specific features in the density that can be calculated relatively easily and allow generalization across systems.  These obviously include the usual semilocal features, but also must include some that capture effects relevant to electron localization, such as the electron number in the vicinity of a nucleus.
Much experimentation is likely to be needed to discover such features.

The next decade should provide widespread use of foundational MLPs, with (we hope) some spectacular scientific progress based on these capabilities.
We can also hope to see XC or orbital-free approximations developed with ML becoming widely available and used, improving in accuracy over human-designed approximations in at least some areas, and maybe even working in new areas where traditional DFT approximations fail.

We shall see.

\section{Funding}
R. A. was supported by JSPS KAKENHI Grant No. 23H04530 from Japan Society for the Promotion of Science (JSPS). M. S. and K. B. are grateful to NSF grant CHE-2154371.

\clearpage

\printbibliography
\end{document}